\documentclass[twocolumn,pra,showpacs,floatfix,preprintnumbers]{revtex4-1}
\usepackage{graphicx}
\usepackage{amsmath}
\usepackage{amssymb}
\usepackage{amsfonts}
\usepackage{upgreek}
\usepackage{txfonts}
\usepackage{color}

\newcommand{\kb}{\mathbf{k}}
\newcommand{\kr}{\mathrm{k}}
\newcommand{\rb}{\mathbf{r}}

\newcommand{\n}{\rm n}
\newcommand{\np}{\rm n^{\prime}}

\newcommand{\kbz}{k_{\rm B}}

\begin{document}

\title{Bose-Einstein Condensation versus Dicke-Hepp-Lieb Transition in an Optical Cavity}
% \title{Finite Temperature Self-Organization \\of a Bose Gas inside an Optical Cavity}

% \author{Francesco Piazza$^{1}$, Philipp Strack$^{2}$, and Wilhelm Zwerger$^{1}$}
\author{Francesco Piazza$^1$}
\email{francesco.piazza@ph.tum.de}
\author{Philipp Strack$^2$}
\author{Wilhelm Zwerger$^1$}
\affiliation{$^{1}$Physik Department, Technische Universit\"at M\"unchen, 85747 Garching, Germany}
\affiliation{$^{2}$ Department of Physics, Harvard University, Cambridge MA 02138}

\begin{abstract}
We provide an exact solution for the interplay between Bose-Einstein condensation
and the Dicke-Hepp-Lieb self-organization transition of an ideal Bose gas trapped inside a single-mode optical
cavity and subject to a transverse laser drive. Based on an effective action approach, we
determine the full phase diagram at arbitrary temperature, which features a bi-critical point 
where the transitions cross. We calculate the dynamically generated band structure of the atoms 
and the associated supression of the critical temperature for Bose-Einstein condensation
in the phase with a spontaneous periodic density modulation. Moreover, we determine the evolution of the polariton spectrum
due to the coupling of the cavity photons and the atomic field near the self-organization transition,
which is quite different above or below the Bose-Einstein condensation temperature. 
At low temperatures, the critical value of the Dicke-Hepp-Lieb transition decreases with temperature and thus thermal fluctuations 
can enhance the tendency to a periodic arrangement of the atoms.
%We observe for instance i) a non-monotonic behavior
%with temperature of the critical coupling for self-organization, ii)
%the existence of a critical temperature $T_0$ for Bose-Condensation which
%depends on the dynamical band structure set by the cavity lattice, and
%iii) broadened polaritonic sidebands drastically changing across $T_0$. 
%
%
%In the thermodynamic limit, and for large atom-pump detunings, our solution of the cavity-Bose gas problem becomes exact. 
%, as in the Dicke model.
%Our approach is based on an effective theory for the single cavity mode
%dressed by the atomic degrees of freedom which becomes exact in the
%thermodynamic limit.
\end{abstract}

%\pacs{PACS numbers}

\maketitle

\section{Introduction}

The study of light-matter interactions at the level of single 
atoms and photons is a major subject in atomic and
molecular physics. The strong atom-field coupling which is 
achievable in high quality cavities has allowed to generate 
entangled states between light and atoms~\cite{haroche_book}
or to directly observe the appearance or disappearance of single photons~\cite{gleyzes_2007}.
The backaction of a single atom onto a cavity can shift its frequency appreciably, as evidenced by a finite vacuum Rabi splitting~\cite{rempe92}.  
%A completely different regime of atoms subject to 
%standing light fields is realized for ultracold atoms in optical lattices \cite{lattice_rmp}.
%In this case, the atoms are bound into the nodes or antinodes of a  
%far-detuned standing light field through the position dependent 
%ac-Stark shift. For sufficiently low atom densities and strong light 
%fields, the lattice due to the periodic modulation of the light intensity 
%is unaffected by the presence of the atoms. A transition between 
%a thermal gas and a Bose-Einstein condensate thus has no 
%effect on the underlying optical lattice. 
New and interesting phenomena at the interface between cavity-QED 
and many-body physics arise in a situation where the the cavity 
contains a finite density of atoms which may undergo either thermal
or quantum phase transitions on their own. A simple case in question
is a cavity filled with an ultracold gas of bosons and subject to
a transverse drive due to the coherent field of a pump laser.  
As predicted by Domokos and Ritsch~\cite{ritsch_2002}, 
the backaction of the atoms on the light in the presence of 
a transverse driving can lead to the formation
of a spontaneous density modulation of the atoms. 
The lattice spacing of the resulting crystalline order is 
set by the cavity mode wavelength, thus optimizing 
the Bragg scattering of the laser photons into the cavity mode.
Simultaneously, the cavity mode itself 
acquires a finite expectation value which leads to
a change of the spectum of polaritons in the cavity. This transition 
to a spatially ordered state of atoms in a cavity has been observed
experimentally first with thermal atoms~\cite{vuletic_2003}
and - more recently - with a Bose-Einstein condensate~\cite{eth_2010,eth_jumps,eth_soft,eth_non_eq}.
These experiments are
conceptually related to the already observed non-steady-state effects
like superradiant Rayleigh scattering
\cite{inouye_1999,kuga_2005,courteille_2007} and collective recoil
lasing \cite{courteille_2007,desalvo_1994}. 

A standard theoretical description of the transition to a self-ordered, periodic 
arrangement of the atoms in a transversely driven cavity assumes that 
the atomic wave vectors can be truncated to just two, namely $\kb=0$ and
the cavity mode wave vector $\kb=\kb_0$.
Within such a two-mode approximation, the problem is reduced to an
effective Dicke model~\cite{dicke_1954} where a large spin of length $J=N/2$
is linearly coupled to the quadrature $\sim\left(\hat{a}^{\dag}+\hat{a}^{}\right)$
of the cavity field. Surprisingly, to date \cite{cavity_rmp}, it has remained unclear how the simplified 
description within a generalized Dicke model \cite{,vukics_2007,carmichael_2007,domokos_2008,larson_2009,domokos_2010,simons_2010, sarang_2010,morigi_2010,domokos_2011,nagy11,bhaseen_2012,oztop12, dallatorre_2012,chang_2013} or the semiclassical treatment of the atomic field 
\cite{domokos_2005,domokos_2006} is connected to the actual many-body situation in which the Bose gas can 
either be Bose-condensed or in a thermal state at higher temperatures. In particular, a study of 
whether and how the properties of the self-organization transition change between the BEC limit studied 
in the ETH experiments~\cite{eth_2010} and the earlier ones in a thermal gas at MIT~\cite{vuletic_2003} 
has been lacking.

\subsection{Key results}
 
Our aim in the present paper is to develop a comprehensive description 
of the backaction between atoms and light field in a driven cavity, which
properly describes the interplay between the Dicke-Hepp-Lieb (DHL) type transition to 
a state with a periodic density modulation of the atoms and the transition
to a phase coherent Bose-Einstein condensate. 
As will be shown below,
the Dicke model provides a correct description of the transition to spatial order of the atoms 
only in the special case of an ideal gas of bosons at zero temperature. Indeed, 
the presence of a continuum of atomic momenta at any finite temperature or,
even at $T=0$, if repulsive interactions are appreciable, precludes a truncation of the 
problem to just two momenta $\kb=0$ and the cavity mode wave vector $\kb=\kb_0$. 

To address the full quantum many-body problem, we develop an effective action approach
that allows to derive the detailed form of the cavity spectrum and the dynamically 
generated bandstructure for the atoms in a fully quantum mechanical 
fashion, without resorting to any semi-classical approximations or truncating the 
Hilbert space to a simplified two-state Dicke model.

Apart from the quite different 
symmetry breakings involved (as we discuss in more detail below), the DHL and BEC transitions are fundamentally distinct 
also in the sense that the former is caused by an external driving in an open
system while the transition to a BEC is due to the standard competition between 
energy and entropy in an equilibrium situation.       
Despite these differences, it will be shown that an effective action approach 
which is based on the assumption of a steady thermal equilibrium state
of the coupled atom-field configuration in a cavity provides a unified 
description of both transitions and even captures some of the 
non-equilibrium features of the DHL transition. Restricting to the case
of non-interacting Bosons in a first step, our main results are:

i) A quantitative phase diagram for the driven cavity problem at
arbitrary temperatures. It contains four different phases which 
are degenerate at a special bi-critical point. The state of the atoms 
in these phases are either a thermal or a BEC phase
which can both be either homogeneous or spatially ordered at 
a wave vector set by the cavity mode and multiples thereof.

ii) A calculation of the dynamical band structure arising for the 
atoms due to the backaction from the cavity field. With 
increasing depth of the temperature dependent 
lattice, the BEC transition temperature is strongly suppressed.

iii) A detailed description of the cavity spectral function associated with polaritonic
excitations which are thermally broadened and sensitive to the
presence or absence of the BEC. In particular, at finite temperature,  the divergence of the emitted
light intensity close to the self-organization transition exhibits a thermal behavior.  

% %%%%%
% {\color{blue}\it PS: what do we mean by ``characteristic singularities''? We should be more specific unless 
% this is standard quantum optics terminology.}
% %%%%%

\subsection{Outline of paper}

In Sec.~\ref{model}, we recapitulate the standard model Hamiltonian and cavity set-up. We also discuss the underlying symmetries 
and the mapping to the Dicke model, to make contact with earlier studies.
In Sec.~\ref{sec:method}, we express the partition function of the coupled atom-cavity set-up in terms of an exact, 
imaginary time path integral. We derive an effective action, nonlocal in space and time, 
which contains the full backaction between the atoms and the cavity field. 
The self-consistent solution of the resulting saddle-point equations allows to determine the full phase diagram, 
the dynamical band structure of the atoms and the behavior of the superfluid condensate and cavity field as function 
of temperature and atom-cavity coupling. The results are discussed in detail in Sec.~\ref{MF}. 
In Sec.~\ref{spectral_function}, we analyze in detail the cavity spectrum, in particular its 
singular behavior near the transition to self-organization and also across the Bose-Einstein condensation temperature. 
In Sec.~\ref{TL}, we discuss the scaling properties of the effective photon action in the limit of large atom numbers.
It is shown that the mean-field solution becomes exact in this limit.
We conclude in Sec.~\ref{sec:conclu} with some open problems and a critical discussion of the advantages and
limitations of using an effective equilibrium formalism for a description of a dissipative open system.

%%%%%%%%%%%%%%%%%%%%%%
\section{Model}
\label{model}

\begin{figure}[t]
\includegraphics[scale=0.4]{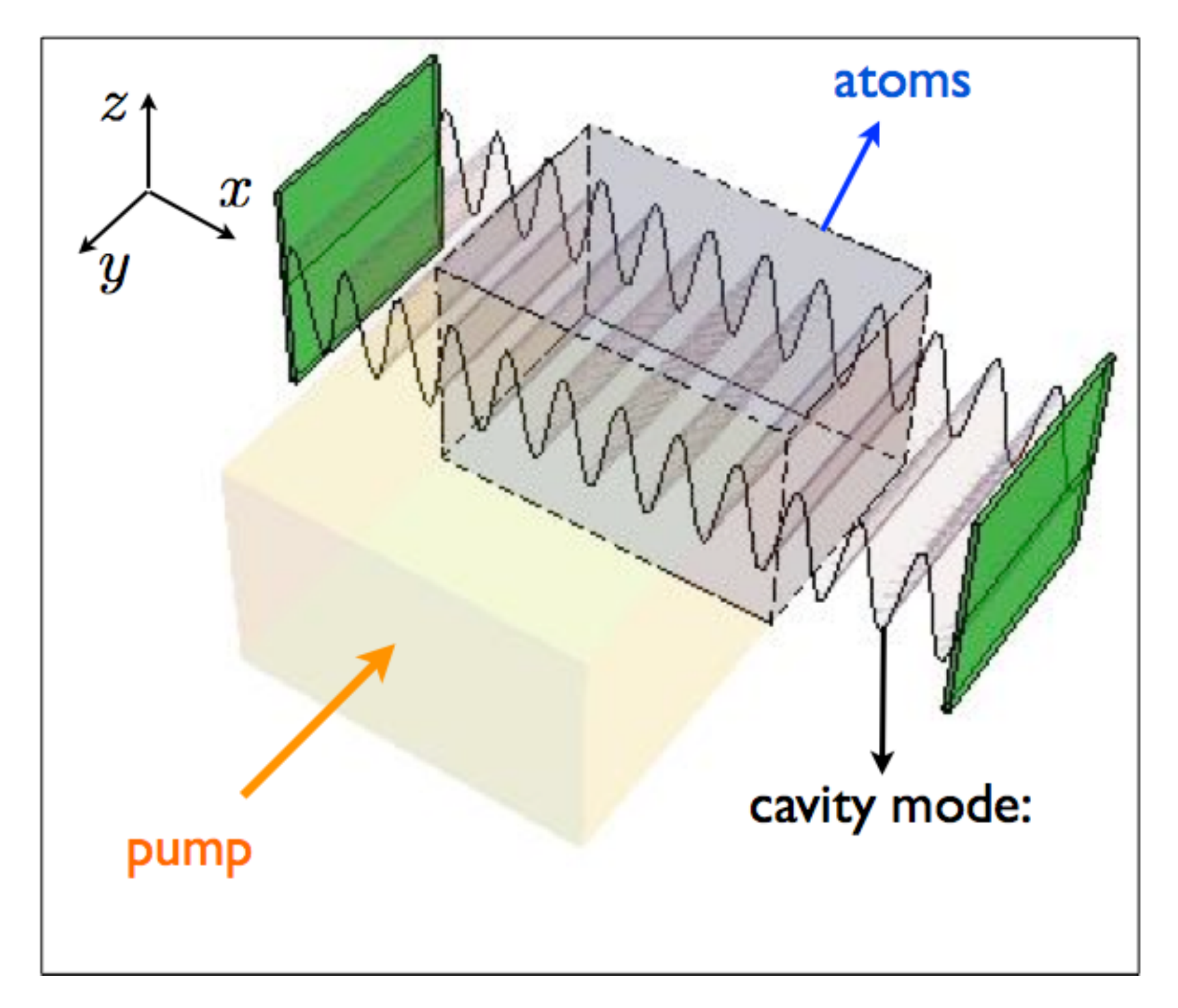}
\caption{The atoms, confined in a three-dimensional box (in blue),
  are interacting with a single cavity mode modulated along the $x$
  direction, and illuminated by the pump laser propagating
  transversally with respect to the cavity axis.}
\label{setup}
\end{figure}

We consider $N$ atoms with two internal electronic levels trapped inside a transversally driven cavity.
The atoms are illuminated by a pump laser and are also coupled
with a single cavity mode. In terms of the quantized field
operators $\hat{\psi}_{g/e}$ for the atoms in the internal ground or excited state
and the annihilation operator $\hat{a}$ for a cavity photon,
the complete atom plus driven cavity Hamiltonian reads~\cite{maschler_2008}
\begin{equation}\label{hamiltonian}
\hat{H}=\hat{H}_{\rm a}+\hat{H}_{\rm c}+\hat{H}_{\rm a/c}+\hat{H}_{\rm
  a/p}\ , 
\end{equation}
where
\begin{align}
\hat{H}_{\rm a}&=\int
d\mathbf{r}\left[\hat{\psi}_g^{\dag}(\mathbf{r})(-\frac{\nabla^2}{2
    m})\hat{\psi}_g^{}(\mathbf{r})+\hat{\psi}_e^{\dag}(\mathbf{r})(-\frac{\nabla^2}{2
    m}-\Delta_{\rm a})\hat{\psi}_e^{}(\mathbf{r})\right]\nonumber\\
\hat{H}_{\rm c}&=-\Delta_{\rm c}\ \hat{a}^{\dag}\hat{a}^{}\nonumber\\
\hat{H}_{\rm a/c}&=-i\ g_0\int d\mathbf{r} \hat{\psi}_g^{\dag}(\mathbf{r})\eta_{\rm c}(\mathbf{r})\hat{a}^{\dag}\hat{\psi}_e^{}(\mathbf{r})+ {\rm h.c}\nonumber\\
\hat{H}_{\rm a/p}&=-i\ \Omega\int d\mathbf{r} \hat{\psi}_g^{\dag}(\mathbf{r})\eta_{\rm p}(\mathbf{r})\hat{\psi}_e^{}(\mathbf{r})+ {\rm h.c}\nonumber
\end{align}
in the frame rotating with the pump frequency
$\omega_{\rm p}$.  Here, $\Delta_{\rm a}=\omega_{\rm p}-\omega_{\rm e}$ and
$\Delta_{\rm c}=\omega_{\rm p}-\omega_{\rm c}$ are the detunings between the pump and
the atomic resonance, and the pump and the cavity mode,
respectively (we set $\hbar=1$ except in some final results). Moreover, $m$ is the atomic mass, $g_0$ is the single-photon Rabi coupling between
the atom and the cavity and $\Omega$ is the pump Rabi frequency. 
The functions $\eta_{\rm c}(\mathbf r), \eta_{\rm p}(\mathbf r)$ contain the
spatial form of the cavity and pump modes, respectively.
The main approximations involved are i) dipole and rotating wave approximation for the
light-matter coupling, ii) neglection of the short-range interaction between atoms
iii) neglection of spontaneous emission from the excited state and of
cavity loss processes through the mirrors. 

\subsection{Symmetry breaking at the Dicke-Hepp-Lieb and the BEC transition}
\label{sec:symmetries}

%We now discuss the two different symmetries of the model Eq.~(\ref{hamiltonian}): (i) the discrete, 
%$Z_2$-symmetry which is spontaneously broken at the DHL transition when the atoms self-organize 
%into one of the two degenerate density patterns, and (ii) the global U(1) gauge symmetry 
%associated with atomic particle conservation that is spontaneously broken when the atoms pick a common 
%phase at the BEC transition. 
In order to discuss which symmetries are broken at either the DHL-transition 
where the atoms acquire a non-vanishing  density modulation and the standard 
BEC transition, we need to fix the shape of the cavity and pump mode profiles. It is also convenient to adiabatically eliminate 
the excited, atomic state. The latter step is justified in the limit of large atom-pump detunings 
$\Delta_a$ and is discussed in more detail below. The resulting, simplified effective Hamiltonian of the 
cavity-Bose gas system that only involves the atomic ground state is \cite{maschler_2008}
\begin{align}
\hat{H}_{\text{eff},a}&=\int d\mathbf{r}\,
\hat{\psi}_g^{\dag}(\mathbf{r})
\left\{
-\frac{\nabla^2}{2m}
+
\frac{\Omega^2}{\Delta_a}
\right\}
\hat{\psi}_g^{}(\mathbf{r})
\nonumber\\
\hat{H}_{\text{eff},c} &= -\Delta_{\rm c}\ \hat{a}^{\dag}\hat{a}^{}
\nonumber\\
\hat{H}_{\text{eff},a/c} &= 
\int d\mathbf{r}\,
\hat{\psi}_g^{\dag}(\mathbf{r})
\Bigg\{\frac{\left(g_0 \eta_c(\mathbf{r})\right)^2}{\Delta_a}\hat{a}^\dagger\hat{a} +\frac{ \Omega g_0\eta_c(\mathbf{r})}{\Delta_a} 
\left(\hat{a} + \hat{a}^\dagger \right)
\Bigg\} 
\hat{\psi}_g^{}(\mathbf{r})\;,
\label{eq:H_eff}
\end{align}
where we have used $\eta_{\rm p}(\rb)=1$, i.e. the spatial structure of 
the pump laser is neglected in accordance with the simplified 
geometry illustrated in Fig.~\ref{setup}. We assume the atoms to be confined in a
cavity with volume $V$ and a cavity mode which extends all over the cloud with
spatial modulation only along the $x$ direction such that 
\begin{align}
\eta_{\rm c}(\mathbf{r})=\cos(\kb_0\cdot\rb)\;\;,\;\;\; \kb_0=(k_0,0,0)\;.
\label{eq:cavity_mode}
\end{align}
The Hamiltonian in Eq.~(\ref{eq:H_eff}) is invariant under a simultaneous parity transformation $\mathcal{P}$
of the photons ($\mathcal{P}: \hat{a}\rightarrow -\hat{a}$) and a
discrete, translational symmetry operation $\mathcal{R}$, which shifts the spatial coordinate of the atoms 
 by odd multiples of the inverse cavity wave vector along the cavity axis direction $x$
 as $\mathcal{R}: \mathbf{r}_x\rightarrow \mathbf{r}_x \pm \pi/k_0$. 
 In terms of a modulated atomic density operator $\hat{\rho}_{\mathbf{k}_0}=\int d\mathbf{r}\,
 \hat{\psi}_g^{\dag}(\mathbf{r})\hat{\psi}_g^{}(\mathbf{r})\cos(\mathbf{k}_0\cdot\mathbf{r})$, the 
 last term in Eq.~(\ref{eq:H_eff}), which is responsible for the self-organization transition, can be written as 
$ \Omega g_0\, \hat{\rho}_{\mathbf{k}_0}\left(\hat{a}+\hat{a}^\dagger\right)$.
The symmetry operations $\mathcal{R}$ and $\mathcal{P}$ act on that term effectively as a $Z_2$
\begin{align}
Z_2:\;\;\;\hat{\rho}_{\mathbf{k}_0}\rightarrow -\hat{\rho}_{\mathbf{k}_0}\;,\;\;\;
%\nonumber\\
\hat{a}\rightarrow -\hat{a}\;,\;\;\;
%\nonumber\\
\hat{a}^\dagger &\rightarrow - \hat{a}^\dagger\; ,
\end{align}
leaving the Hamiltonian invariant under the product  $\mathcal{R}\mathcal{P}$.
 At the self-organization transition, the cavity develops 
a coherent field $\pm\langle \hat{a} + \hat{a}^\dagger \rangle\ne 0$ into one of the two parity states available.
In contrast to the BEC transition, there is no continuous symmetry breaking, however, 
associated with the appearance of condensate in the cavity field.  Indeed the phase of the coherent state is
locked to the phase of the pump laser with an extensive coupling energy $E_J^{\rm th}$, see Eq. (27) below.

Simultaneous with the appearance of a non-vanishing average cavity field, the atoms 
spontaneously arrange into one of the two available checkerboard patterns 
$\pm\langle \hat{\rho}_{\mathbf{k}_0}\rangle$. As will be discussed in Section V, there is
a soft mode associated with the formation of the lattice structure in the atoms, 
whose frequency vanishes at the transition. Due to the 
discrete symmetry breaking, this is not a Goldstone mode, however, and - moreover -
it is overdamped near the transition (see Eq. (45) below).  

The second symmetry of Eq.~(\ref{eq:H_eff}) is the continuous $U(1)$-symmetry
associated with the conservation of the number of atoms.  Eq.~(\ref{eq:H_eff}) is invariant under the global shift
\begin{align}
U(1): \theta(\mathbf{r})\rightarrow \theta(\mathbf{r})+ c\;,
\label{eq:phase}
\end{align}
by a constant, real-valued number $c$ of the phase of the atomic field operator
$\hat{\psi}_g (\mathbf{r}) = \rho(\mathbf{r}) e^{i\theta(\mathbf{r})}$. While this 
symmetry is not spontaneously broken in the relevant case with a fixed number 
of atoms in the cavity, the atomic field still exhibits off-diagonal long range order
below the BEC transition temperature. The associated non-vanishing
rigidity energy $\sim \left(\nabla\, \theta(\mathbf{r})\right)^2$ with respect to small gradients 
of the phase implies the existence of a proper Goldstone mode for the BEC 
transition~\cite{lattice_rmp}. In the non-interacting gas, its dispersion is that of free particles $\epsilon(\kb)=\hbar^2\kb^2/2m$,
while for any finite, repulsive interaction it is a linearly dispersing Bogoliubov sound mode.
The change from a Goldstone mode with quadratic to one with a linear dispersion
upon introducing interactions is due to the finite compressibility in the latter case, 
which entails a formal relativistic invariance in the quantum hydrodynamic 
description of the broken symmetry phase~\cite{lattice_rmp}. 

%More generally,
%non-relativistic microscopic Hamiltonians can lead to Goldstone modes of
%both types, see e.g. T. Schaefer et. al Phys. Lett. B522, 67 (2001) and A. Kapustin, arXiv:1207.0457. 
% %%%%%
 %{\color{blue}\it WZ: The last sentence may be more for our own interest}
% %%%%%

\subsection{Hilbert space truncation to $Z_2$ Dicke model}

As mentioned in the Introduction and discussed in detail in Refs.~\onlinecite{eth_2010,domokos_2010}, 
one can map Eq.~(\ref{eq:H_eff}) to the Dicke model upon truncating the Hilbert space of the 
atoms to containing two momentum states only. 
Within such a two-mode approximation, the problem is reduced to an
effective Dicke Hamiltonian~\cite{dicke_1954} 
\begin{equation}
\hat{H}=\delta_{\rm c}\
\hat{a}^{\dag}\hat{a}^{}+E_{\rm R}\hat{J}_{\rm
  z}+\frac{\lambda}{\sqrt{2N}}\left(\hat{a}^{\dag}+\hat{a}^{}\right)\left(\hat{J}_{\rm +}+\hat{J}_{\rm -}\right)\;,
\label{eq:effective_dicke}
\end{equation}
where a large spin of length $J=N/2$ is linearly coupled to the quadrature
of the cavity field. The collective angular momentum operators in 
the two-mode Hilbert space of the $N$ atoms are: $\hat{J}_{\rm
 z}=\frac12\sum_{j=1}^N(|\mathbf{k_0}\rangle_j\langle
\mathbf{k_0}|-|\mathbf{0}\rangle_j\langle \mathbf{0}|)$, $\hat{J}_{\rm
 +}=\sum_{j=1}^N|\mathbf{k_0}\rangle_j\langle
\mathbf{0}|$ and $\hat{J}_{\rm
 -}=\sum_{j=1}^N|\mathbf{0}\rangle_j\langle \mathbf{\kb_0}|$.
The effective magnetic field which couples to the spin polarization
$\hat{J}_{\rm z}$ is determined by the recoil energy $E_{\rm R}=\epsilon(\kb_0)=\hbar^2\kb_0^2/(2m)$ 
of a single atom at the wave vector $\kb_0$ set by the cavity  field.  
The Rabi coupling $\lambda$, 
the effective detuning of the cavity mode with respect to the atomic resonance $\delta_c$,  and the single-atom dispersive shift $u_0$
\begin{equation}
\lambda=\frac{\Omega g_0}{\Delta_{\rm a}}\sqrt{N}\;,\;\;\;\;\delta_{\rm c}=-\Delta_{\rm c}+\frac12 u_0\;,\;\;\;\;\;u_0=\frac{g_0^2}{\Delta_{\rm  a}}N
\end{equation}
are all defined so that they approach finite values $\propto \sqrt{n}$ in the thermodynamic limit (TL): $N,V\to\infty$,
$N/V=n=\text{const}$. Indeed, the
bare coupling $g_0$ decreases like $1/\sqrt{V}$ when the cavity volume is
increased. Typical order of magnitudes for these couplings in quantum-optical experiments are recoil energies
$E_{R}$ in the kHz regime and 
pump-cavity detunings $\Delta_c$ in the MHz regimes, while temperatures $T$ are in the micro-Kelvin
to nano-Kelvin regime \cite{cavity_rmp}.

We note that in the effective Hamiltonian (\ref{eq:effective_dicke})
the system volume $V$ does not play any role once we know that the
effective parameters are intensive quantities. In the Dicke model
only the number of two-level atoms $N$ enters and the TL is simply $N\to\infty$.
In this limit, a standard Holstein-Primakoff expansion 
$J_{-}=\sqrt{N-\hat{b}^\dag\hat{b}}\hat{b}\simeq\sqrt{N}\hat{b}$
of the angular momentum operators reduces the Dicke model to a rather simple Hamiltonian of two
harmonic oscillators with a linear coupling $\sim (\hat{a}^\dag+\hat{a})(\hat{b}^\dag+\hat{b})$, which 
leads to a crossing of the two eigenfrequencies \cite{emary03}. 

Eq.~(\ref{eq:effective_dicke}) maps the Bose gas in a cavity to 
a zero-dimensional problem of a single, large$-N$ spin coupled 
to a single harmonic oscillator. The 
two degenerate density patterns of the atoms discussed above
correspond to the two degenerate orientations of a large Ising spin 
with length proportional to the number of atoms $N$. Note that Eq.~(\ref{eq:effective_dicke}) 
contains counter-rotating terms and is invariant under a discrete $Z_2$ transformation:
%
%\begin{align}
$\hat{J}_+\rightarrow -\hat{J}_+
%\nonumber\\
\;,\;\;
\hat{J}_-\rightarrow -\hat{J}_-
%\nonumber\\
\;,\;\;
\hat{a}\rightarrow -\hat{a}
%\nonumber\\
\;,\;\;
\hat{a}^\dagger \rightarrow - \hat{a}^\dagger\;.
$
%\end{align}
%
Without counter-rotating terms the so-called $U(1)$ Dicke model has an additional global $U(1)$ symmetry 
\cite{emary03,ye_2011} which is not to be confused with the $U(1)$ of the atoms introduced in Eq.~(\ref{eq:phase}).

The Dicke model has a number of important properties which - later on in our paper - we will compare
with the results for the full problem which does not rely on the two-state truncation:

(a) {\it Exact solvability}: The Dicke model \cite{dicke_1954} is a zero-dimensional model and can be solved exactly at arbitrary 
temperatures in the TL $N\rightarrow \infty$ \cite{lieb_1973, hioe_1973}. 
It exhibits the Dicke-Hepp-Lieb (DHL) transition to a superradiant state, in which the ground state has a finite occupation of the 
photon mode combined with a finite atomic polarization. 

(b) {\it Quantum bifurcation rather than phase transition}: The DHL transition is a quantum bifurcation in a 
zero-dimensional system. As a result one may define a 
critical exponent $z\nu=1/2$~\cite{emary_2004} from
the vanishing of the soft mode frequency but the 
exponents  $z$ and $\nu$ have no seperate meaning 
because there is no divergent correlation length here.

(c) {\it Finite temperature phase boundary}: At finite temperature,
the critical spin-photon coupling turns out to be~\cite{lieb_1973,hioe_1973}
\begin{equation}
\label{threshold_DM}
\lambda_{\rm D}^2(T)=\frac{E_{\rm R}\delta_{\rm
    c}}{2\tanh\left(E_{\rm R}/2\kbz T\right)}\;,
\end{equation}  
As will be shown in Sec.~\ref{so_th}, the Dicke model gives the correct value for the critical coupling both at 
very low and at high temperatures $k_BT\gg E_R$. However, within the full theory, 
there is an intermediate regime, at which thermal fluctuations in fact favor a spatially ordered 
pattern of the atoms. This is not captured within the Dicke model expression for the critical coupling 
Eq.~(\ref{threshold_DM}), which is instead a monotonously growing
function of temperature $T$. Moreover, the temperature dependence
of the number of condensed atoms, also not included in the Dicke model, affects the critical coupling 
in a non-trivial way.

\section{Effective Action Approach}
\label{sec:method}

In this section, we derive a Landau-Ginzburg-Wilson type effective action
for the coupled order parameters which characterize both the 
DHL- and the BEC transition. Beyond the complex cavity field, the order parameter 
of the atoms is a spinor with an infinite number of terms associated with Fourier components
of the condensate at wave vectors which are an arbitrary multiple of $\kb_0$.
The effective action is derived by eliminating all
other degrees of freedom from our problem, which can be conveniently done using the path-integral formalism.
Starting with the time-independent Hamiltonian (\ref{hamiltonian}), the associated partition function can be
expressed as a functional integral in the form 
\[
\mathcal{Z}=\int D(\psi_e^*,\psi_e^{\phantom{*}}) D(\psi_g^*,\psi_g^{\phantom{*}}) D(a^*,a\!\!^{\phantom{*}})e^{-S[\psi_e^*,\psi_e^{\phantom{*}} \psi_g^*,\psi_g^{\phantom{*}},a^*,a^{\phantom{*}}\!\!]}.
\]
We decompose the fields as
\[
\psi_{g/e}(\mathbf{r},\tau)=\frac{1}{\beta\sqrt{V}}\sum_{\n,\kb}e^{-i\omega_{\n}\tau+i\kb\cdot\mathbf{r}}\psi_{g/e;\n\kb}\
,\ 
a(\tau)=\frac{1}{\beta}\sum_{\n}e^{-i\omega_{\n}\tau}a_{\n}
\] 
into the Fourier components 
of the fluctuating atom and cavity field in imaginary time $\tau$.
Here $\omega_{\n}=2\pi\n/\beta$ is a bosonic Matsubara frequency and
$\beta=1/(\kbz T)$ the inverse temperature. The different contributions to the 
Hamiltonian (\ref{hamiltonian}) give rise to a corresponding sum of action terms
\begin{equation}\label{action_start}
S\left[\psi_e^*,\psi_e^{\phantom{*}} \psi_g^*,\psi_g^{\phantom{*}},a^*,a^{\phantom{*}}\!\!\right]=S_{\rm a}+S_{\rm c}+S_{\rm a/c}+S_{\rm
  a/p}\ , 
\end{equation}
which read
\begin{align}
S_{\rm
  a}&=\frac1\beta\sum_{\n,\kb}\psi_{e;\n\kb}^{*}\left(-i\omega_{\n}+\epsilon(\kb)-\mu-\Delta_{\rm
    a}\right) \psi_{e;\n\kb}^{\phantom{*}}\nonumber\\ &+\frac1\beta\sum_{\n,\kb}\psi_{g;\n\kb}^{*}\left(-i\omega_{\n}+\epsilon(\kb)-\mu\right) \psi_{g;\n\kb}^{\phantom{*}}\nonumber\\
S_{\rm
  c}&=\frac1\beta\sum_{\n,\kb}a_{\n}^{*}\left(-i\omega_{\n}-\Delta_{\rm
    c}\right) a_{\n}^{\phantom{*}}\nonumber\\
S_{\rm
  a/c}&=\frac{ig_0}{2\beta^2}\!\!\!\sum_{\n,\np,\kb}\!\!\!
a_{\n-\np}^{*}\psi_{g;\np\kb}^{*}\left(\psi_{e;\n\kb-\kb_0}^{\phantom{*}}+\psi_{e;\n\kb+\kb_0}^{\phantom{*}}\right)+{\rm h.c}\nonumber\\
S_{\rm
  a/p}&=\frac{i\Omega}{\beta}\sum_{\n,\kb}\psi_{g;\n\kb}^{*}\psi_{e;\n\kb}^{\phantom{*}}+
{\rm h.c}\ ,\nonumber
\end{align}
where $\epsilon(\kb)=\hbar^2 k^2/2m$ and $k=|\kb|$. 

As a first step, we integrate out the atomic field associated with the excited atomic level. Since the
above action is quadratic in $\psi_{e;\n\kb}$ plus source terms which
are linear in $\psi_{e;\n\kb}$, this elimination can be performed exactly by evaluating a 
Gaussian path-integral. 
% using the known result: $\int
% D(z^*,z)\exp\left(-z^* A
%   z+z^*b+b^*z\right)=\det(A)^{-1}\exp\left(b^*A^{-1}b\right)$.
For an arbitrary value of the detuning $\Delta_a$, this introduces retardation effects 
with complicated effective couplings between the ground state atoms
and the cavity field. The situation is considerably simplified in the 
experimentally relevant limit, where the detuning $\Delta_a$,
typically in the {\rm GHz} regime, is much larger than 
the recoil energy and the atomic chemical potential, typically in the {\rm KHz} regime.
In this limit, the dynamics of the atoms in the excited
states is irrelevant and the propagator  $(-i\omega_{\n}+\epsilon(\kb)-\mu-\Delta_{\rm a})^{-1}$
for the excited atoms can be replaced by a constant $1/\Delta_{\rm a}$.  
As a result, the excited atomic level is eliminated adiabatically,  
yielding 
\begin{align}
S^{\prime}\left[\psi^*,\psi^{\phantom{*}}\!\!,a^*,a^{\phantom{*}}\!\!\right]=\frac1{\beta^2}\sum_{\rm
  n,m}\sum_{\mathbf{k}\in B}\Psi_{\rm
  n}^{\dag}(\mathbf{k})\mathrm{M}_{\rm n,m}(\mathbf{k})\Psi_{\rm
  m}^{}(\mathbf{k})\nonumber\\+\frac1\beta\sum_{\rm n}(-i\omega_{\rm
  n}-\Delta_{\rm c})|a_{\rm n}|^2\ .
\label{S_prime}
\end{align}
as an effective action for the
ground state atoms plus the cavity field (in the following, we will drop the subscript
$g$ indicating the ground state atoms).
Here, we employed a Nambu spinor representation 
\begin{align}
&\Psi_{\rm n}^T(\kb)=\nonumber\\&\left(
\begin{array}{ccccccc}
\dots&\psi_{{\rm n},\kb-2\kb_0}&\psi_{{\rm n},\kb-\kb_0}&\psi_{{\rm n},\kb}&\psi_{{\rm n},\kb+\kb_0}&\psi_{{\rm n},\kb+2\kb_0}&\dots
\end{array}
\right)\ ,
\end{align}
for the atomic field, where every component of the spinor corresponds to a different Bloch
band set by the cavity mode. The momentum sum in the
action (\ref{S_prime}) is therefore restricted to the first Brillouin
zone $\kb\in B$,  i. e. $\kb=(-k_0/2<\kr_x<k_0/2,-\infty<\kr_y<\infty,-\infty<\kr_z<\infty)$ in the 
simple geometry shown in Fig.~\ref{setup}. 
This spinor representation has the advantage that the matrix $\mathrm{M}_{\rm n,m}(\mathbf{k})$
in Eq.~(\ref{S_prime}), containing the coupling between cavity and
atoms, is diagonal in quasi-momentum $\kb$. Its explicit form
\begin{widetext}
\begin{align}\label{matrix_full} 
\mathrm{M}_{\rm n,m}(\mathbf{k})=
\left(
\begin{array}{ccccccc}
\dots & \dots & \dots & \dots & \dots & \dots & \dots \\
U_{\rm n,m}/2 & \Lambda_{\rm n,m}/2 &
d_{\rm n,m}(\mathbf{k}-\mathbf{k_0})&
\Lambda_{\rm n,m}/2   &   U_{\rm n,m}/2&
 0 & 0\\
0 & U_{\rm n,m}/2 & \Lambda_{\rm n,m}/2 &
d_{\rm n,m}(\mathbf{k})&
\Lambda_{\rm n,m}/2  & U_{\rm n,m}/2&
 0 \\
0 & 0 & U_{\rm n,m}/2 & \Lambda_{\rm n,m}/2 &
d_{\rm n,m}(\mathbf{k}+\mathbf{k_0})&
\Lambda_{\rm n,m}/2   &  U_{\rm n,m}/2 \\ 
\dots & \dots & \dots & \dots & \dots & \dots & \dots
\end{array} 
\right)\ .
\end{align}
\end{widetext}
is tridiagonal in Nambu space. Its diagonal elements 
\begin{equation}
\label{diag_el}
d_{\rm n,m}(\mathbf{k})=\delta_{\rm
  n,m}G_0^{-1}(\omega_{\n};\kb)+U_{\rm n,m}
\end{equation}
contain the inverse free boson propagator $G_0^{-1}(\omega_{\n};\kb)=\beta(-i\omega_{\n}+\epsilon(\kb)-\mu)$,
where $\mu$ is the chemical potential, which is shifted compared to
its bare value by a contribution $\Omega^2/\Delta_{\rm a}$ due to the
spatially homogeneous external drive. 
The propagator $G_0$ describes the
intra-band dynamics of the atoms. 

\begin{figure}[b]
\includegraphics[scale=0.3]{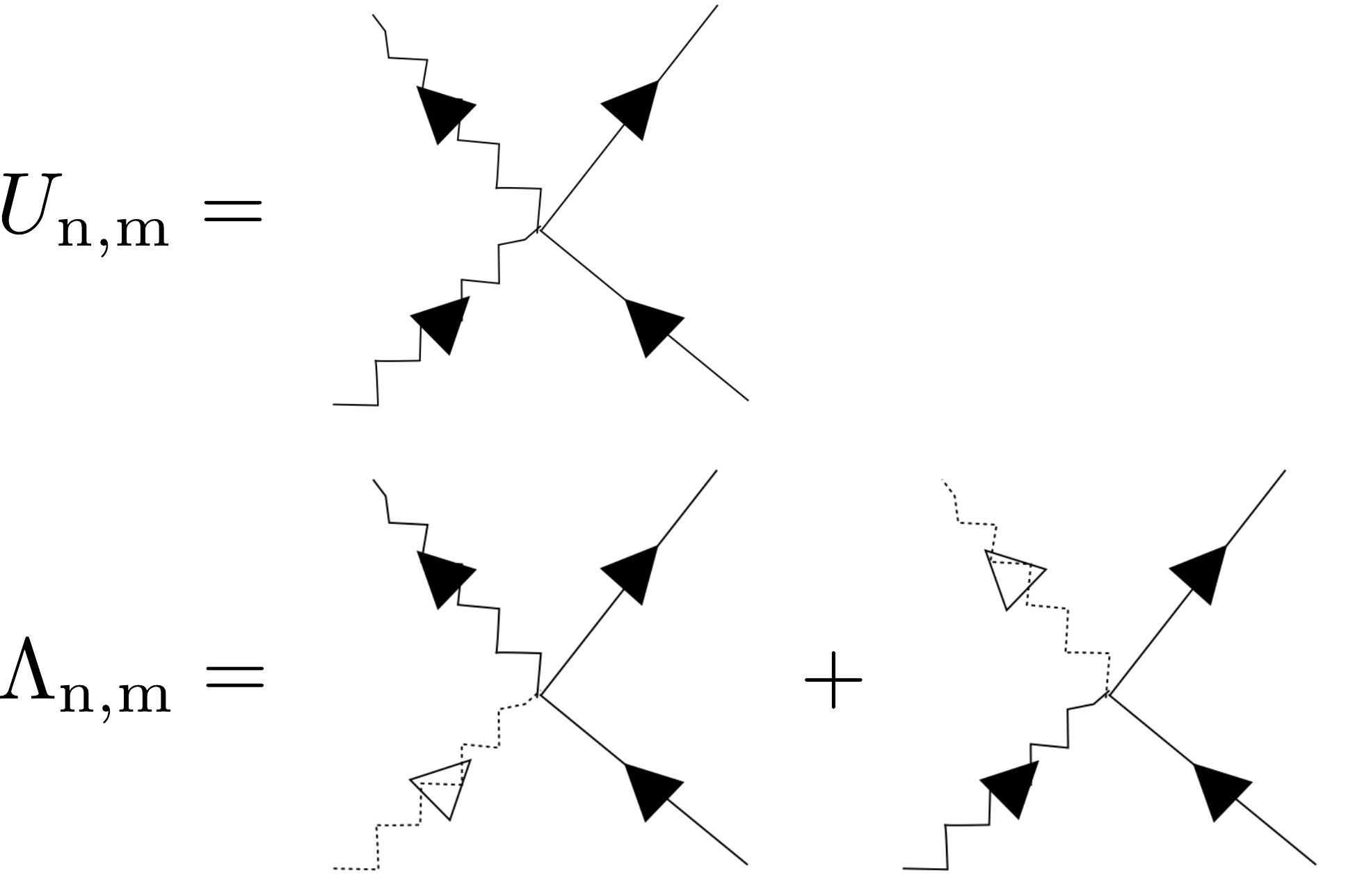}
\caption{Feynman diagrams for the effective scattering vertices
  between ground state atoms and photons. Solid lines correspond to
  atoms, zig-zag lines to cavity photons, dashed lines with empty
  arrows to pump photons.}
\label{effective_vertices}
\end{figure}

The second term
\begin{equation}
\label{eq_Unm}
 U_{\rm n,m}=\frac{u_0}{2\beta N}\sum_{\rm n_1}a_{\rm n_1-n}^{*}a_{\rm
   n1-m}^{\phantom \star}\ ,
\end{equation}
 in Eq.~(\ref{diag_el}) results from two photon processes where the
 atom absorbs a photon from the cavity and re-emits it into the
 latter. The corresponding effective vertex after elimination of the
 excited atomic level is illustrated by the upper Feynman diagram in Fig.~\ref{effective_vertices}. 
The atom-cavity coupling described by Eq. (\ref{eq_Unm}) is responsible for the dispersive shift of
the cavity frequency induced by the atoms. More precisely, the shift is
 described by the diagonal part in Matsubara space ${\rm n=m}$, while the
off-diagonal part appears since the atoms can
also induce fluctuations in the cavity and vice versa.

The off- diagonal elements of the matrix (\ref{matrix_full}) describe
the coupling between nearest and next-to-nearest neighbouring Bloch bands.
The cavity field $a(\tau)$ only depends on the imaginary time 
variable $\tau$ since its spatial form is assumed to be always given by
the stationary mode eigenfunction $\eta_{\rm c}=\cos(\kb_0\cdot\rb)$. 
In line with most previous work, we keep the spatial structure of the photon mode 
unchanged by the presence of atoms or potential optomechanical couplings 
between the mirrors and intracavity photon field. This is justified as
long as the dipole coupling $g_0$ is much smaller than the energy
separation between the cavity modes. This ensures the effective 
zero-dimensionality of the problem and is crucial for the exact
solvability of the model, as we discuss in Sec.~\ref{TL}. Including dispersive effects of the photons 
or self-consistently determining the cavity mode functions as a function of momentum 
significantly complicates the analysis and we leave this to future work.

In the Feynman diagrams in momentum space, every cavity photon line thus has two 
possible momenta $\pm\kb_0$, corresponding to photons travelling in opposite
directions. This is why we have both super- and sub-diagonals in the
matrix (\ref{matrix_full}).
For the same reason, the diagram for $U_{\rm n,m}$ can contribute
to both the intra-band coupling, when the atom absorbs a
photon travelling in one direction and emits it in
the same direction, and also to the next-to-nearest neighbor coupling,
when the absorbed and emitted photons have opposite directions.

The presence of a transverse pump laser finally leads to a coupling 
\begin{align}
\label{lambda_nm}
\Lambda_{\rm n,m}=\frac{\lambda}{\sqrt{N}}(a_{\rm m-n}^{*}+a_{\rm n-m}^{\phantom *})\ ,
\end{align}
between nearest neighboring bands.
Note that the effective cavity-atom coupling is actually a two-photon process proportional to the amplitude 
of the pump laser $\Omega$. So even for relatively small dipole matrix elements $g_0$, by increasing the pump amplitude
$\Omega$, one can achieve sufficiently strong couplings $\lambda$ to induce self-organization. The Feynman diagrams corresponding to Eq.~(\ref{lambda_nm}) are shown
in Fig.~\ref{effective_vertices}.
In this effective vertex, the cavity photon is turned into a pump photon
upon scattering with a ground state atom (or the other way
around). This corresponds to an atom absorbing a photon from the pump
and emitting it into the cavity (or the other way around).

In a second step towards an effective action which only contains the 
order parameters for both the DHL and BEC transition, we integrate out the 
fluctuations in the ground state atomic field $\Psi_{\n}(\kb)$. In order to account 
for the possible occurrence of BEC, we separate out the condensate part in the standard form
\begin{equation}
\Psi_{\rm n}^{}(\mathbf{k})=\beta\delta_{\rm
  n,0}\delta_{\mathbf{k},0}\sqrt{N}\Phi_0+\delta\Psi_{\rm n}^{}(\mathbf{k})\ ,
\label{separ}
\end{equation}
where
\begin{align}\label{cond}
\Phi_0=\left(
\begin{array}{ccccccc}
\dots&\phi_{-2\kb_0}&\phi_{-\kb_0}&\phi_0&\phi_{\kb_0}&\phi_{2\kb_0}&\dots
\end{array}
\right) 
\end{align}
is the spinor describing the condensate wavefunction in Nambu
space. It defines the number of atoms in the condensate $N_0=N\Phi_0^\dag\Phi_0$.
At this point, it is important to realize that the BEC phase 
discussed here is of a multi-mode character in the regime where
the cavity field acquires a non-vanishing average occupation. 
Indeed, as discussed before in the special case of  zero temperature~\cite{domokos_2011},
a macroscopic occupation of momentum states of the atoms then appears not only in the $\kb =0$ mode but also
in harmonics $\pm n\kb_0$ of arbitrary order $n=1,2,\dots$ in the 
cavity wavevector $\kb_0$. 

Since the scattering with a cavity photon
cannot change the quasi-momentum of the atom but only its Bloch band, only the finite
temperature can induce an occupation of a quasi-momentum different
from zero. Therefore, as long as short-range interactions between
the atoms are excluded, it is completely general to assume that the
condensate spinor can only have quasi-momentum equal to zero, as done
in Eq.~(\ref{cond}).
Moreover, in the absence of short-range interactions, the action (\ref{S_prime}) contains terms 
only up to second order in
$\delta\Psi_{\rm n}^{}(\mathbf{k})$. The fluctuations of the ground state atoms 
can thus again be eliminated exactly by a 
Gaussian integral. By accounting properly for the non-standard terms 
which are linear in $\delta\Psi_{\rm n}^{}(\mathbf{k})$ and contain 
the condensate spinor $\Phi_0$, the resulting effective action is finally given by 
\begin{align}\label{S_eff}
&S_{\rm eff}[\Phi_0^\dag,\Phi_0^{\phantom{\dag}},a^*\!,a]=\frac1\beta\sum_{\rm n}(-i\omega_{\rm n}-\Delta_{\rm c})|a_{\rm n}|^2+\mathrm{Tr}\ln\left(\mathrm{M}\right)+\nonumber\\
&N\Phi_0^{\dag}\mathrm{M}_{\rm 0,0}(\mathbf{0})\Phi_0-N\!\!\!\sum_{\rm
  n,m\neq 0}\!\!\!\Phi_0^\dag\mathrm{M}_{\rm 0,n}(\mathbf{0})\mathrm{M}_{\rm n,m}^{-1}(\mathbf{0})\mathrm{M}_{\rm
  m,0}(\mathbf{0})\Phi_0\ .
\end{align}
In the tracelog term, the trace corresponds to
$\sum_{\n}\sum_{\kb\in B}\mathrm{tr}$, where
$\mathrm{tr}$ indicates the trace in Nambu space. This
term describes the contribution to the energy of the cavity photons
originating from the coupling with thermal atoms only. % It is extensive,
% as is the first term describing the bare cavity field. 
The terms on the second line of Eq.~(\ref{S_eff})
describe the contribution to the energy of the cavity photons due to
the presence of condensed atoms. In particular, the last term describes
processes where both condensed and thermal atoms are involved.
We point out that here the summation is
restricted to ${\rm n,m}\neq 0$. This restriction comes from requiring that the
atomic fluctuations be distinct from the
condensate part in the definition (\ref{separ}). More formally, one
requires the fluctuations field $\delta\Psi_{\n}(\kb)$ to have no
$\n=0,\kb=0$ part. The exclusion of the $\kb=0,{\rm n,m}=0$ part is superfluous in
the tracelog term since there we have an additional sum over
$\kb$. The $\kb=0$ part would have thus no weight in the thermodynamic
limit where the sum becomes an integral.

The action (\ref{S_eff}) is still exact, since the elimination of
the atomic degrees of freedom could be performed exactly. Note, however, that the action contains arbitrary 
orders in the cavity field $a_{\rm n}$ through the logarithm and the
inverse of the matrix $M_{\rm n,m}(\kb)$.
The nontrivial second order contribution describes how a cavity photon gets 
dressed by the atoms. The higher order terms describe the interactions
between photons mediated by the atoms.

% Having derived the effective action for the cavity photons, the first
% step will be to perform a mean-field (MF) analysis by solving the classical equations of motion
% stemming from the action (\ref{S_eff}).
% This will yield the finite temperature phase diagram discussed in section \ref{MF}.
% In section \ref{spectral_function}, we will then study the fluctuations about the MF by expanding the
% action in powers of the cavity field. This will allow us to calculate
% the photon self-energy, which tells us how the photons gets dressed by
% the atoms. Subsequently, we will analyze the behavior of the cavity
% fluctuations in the thermodynamic limit (TL), which has important
% consequences on the nature of the self-organization transition.
% Finally, we will study the spectrum of the cavity fluctuations by
% means of the spectral function..

\subsection{Saddle-point equations}

In the following, we will determine the complete phase diagram 
within a mean-field (MF) treatment.
In our present problem, where direct interactions between the atoms are neglected, this
saddle point approximation is in fact exact in the TL.% limit $N\to\infty$ because an overall
% factor $N$ can be pulled out from the effective action (\ref{S_eff}.)
This will be discussed in detail in section \ref{TL} below. To determine the phase diagram
within MF,  one needs to find non-vanishing solutions of the
equation(s) which correspond to extrema of the action. Specifically, 
we have to minimize with respect to both the cavity and condensed atoms
fields
\begin{align}\label{extrem}
\frac{\partial S_{\rm eff}[\Phi_0^\dag,\Phi_0^{\phantom{\dag}},a^*\!,a]}{\partial (a_{\rm n}^*,\partial \Phi_0^\dag)}&=0
% \nonumber\\
% %
% \frac{\partial S_{\rm eff}[\Phi_0^\dag,\Phi_0^{\phantom{\dag}},a^*\!,a]}{\partial \Phi_0^\dag}&=0
% \nonumber\\
% %
% \frac{1}{\beta}\frac{\partial S_{\rm eff}[\Phi_0^\dag,\Phi_0^{\phantom{\dag}},a^*\!,a]}{\partial \mu }&=N\;,
\end{align}
In addition, we need to satisfy the equation of state 
\begin{equation}
\label{eq:eos}
N=\frac{1}{\mathcal{Z}}\left(\frac{\partial\mathcal{Z}}{\beta\partial_\mu}\right)_{T,V}\ ,
\end{equation}
which determines the average atom density $n=N/V$ as a function of the chemical potential $\mu$.

%Since after adiabatic
%elimination of the excited atomic level our effective description is
%invariant under time translation, 
Both the cavity and the atom field are indepent of the imaginary time $\tau$
in the  MF solution. The MF cavity field thus has only the zero
frequency Matsubara component: 
\begin{equation}
a_{\n}^{\rm MF}=\beta\delta_{\rm n,0}\sqrt{N}\alpha\ .
\end{equation}
%As we shall show in Section \ref{TL}, this saddle-point stationary field
%approximation becomes exact in the TL. Within this framework,
Below the DHL transition the state of the cavity is the vacuum
$\alpha=0$, while the cavity is in a coherent
state with amplitude equal to $\sqrt{N}\alpha$ above the transition. 

In order to determine the full phase diagram, the values of both $\alpha$ and $\Phi_0$ have to be
calculated by solving the system of two coupled equations
\begin{equation}\label{mf_eqs}
\begin{cases}
\ \mathcal{M}_{\alpha}(0;\mathbf{0})\Phi_0=0  \\
\ -\Delta_{\rm
  c}\alpha+\Phi_0^\dag\mathcal{M}_{\alpha}^{\prime}\Phi_0+\frac{1}{n}\displaystyle\sum_{\rm
  n}^{\phantom{n}}\int_{\mathcal{B}}\frac{d\kb}{(2\pi)^3}\mathrm{tr}\left[\mathcal{M}_{\alpha}^{-1}(\omega_{\rm
  n};\kb) \mathcal{M}_{\alpha}^{\prime}\right]=0  
\end{cases}
\end{equation}
which arise from the saddle point condition (\ref{extrem}). In addtion, 
we have to satisfy the equation of state 
\begin{equation}\label{mf_eos}
\Phi_0^\dag \Phi_0^{\phantom{\dag}}+\frac{1}{n}\displaystyle\sum_{\rm
  n}^{\phantom{n}}\int_{\mathcal{B}}^{\phantom{B}}\frac{d\kb}{(2\pi)^3}\mathrm{tr}\left[\mathcal{M}_{\alpha}^{-1}(\omega_{\rm
  n};\kb) \right]=1 \ ,
\end{equation}
obtained from (\ref{eq:eos}) by evaluating the partition function in
the saddle point approximation $\mathcal{Z}\simeq \exp(-S_{\rm eff}^{\rm MF})$.
After solving these equations, the resulting MF action can be written as $S_{\rm eff}^{\rm MF}=\beta
 F_{\rm eff}^{\rm MF}$ where
\begin{align}\label{s_mf}
F_{\rm eff}^{\rm MF}=-N\ \Delta_{\rm c}|\alpha|^2+\frac{V}{\beta}\sum_{\rm
  n}\int_{\mathcal{B}}^{\phantom{B}}\frac{d\kb}{(2\pi)^3}\mathrm{tr}\ln\left[\mathcal{M}_\alpha (\omega_{\rm
  n};\kb)\right]\nonumber\\+ N\ \Phi_0^{\dag}\frac{\mathcal{M}_\alpha(0;\mathbf{0})}{\beta}\Phi_0
\end{align}
is the free energy, which is a minimum for the given values of $\Phi_0$ and $\alpha$.

In Eqs.~(\ref{mf_eqs}),(\ref{mf_eos}) and (\ref{s_mf}), we substituted the momentum sum with an integral: $\sum_{\kb\in
  B}\to V\int_{\mathcal{B}} d\kb/(2\pi)^3$,
and defined the MF Nambu matrix 
\begin{widetext}
\begin{align} 
\frac1\beta\mathcal{M}_{\alpha}(\omega_{\rm
  n};\kb)=\mkern410mu\nonumber\\
\left( 
\begin{array}{ccccccc}
\dots & \dots & \dots & \dots & \dots & \dots & \dots  \\
 \frac{u_0}{4}|\alpha|^2 & \frac{\lambda}{2}(\alpha^*+\alpha) &
 \frac1\beta G_0 ^{-1} (\omega_{\n};\kb-\kb_0)+\frac{u_0}{2}|\alpha|^2 &
 \frac{\lambda}{2}(\alpha^*+\alpha)  &  \frac{u_0}{4}|\alpha|^2 &
 0 & 0\\
0 & \frac{u_0}{4}|\alpha|^2 & \frac{\lambda}{2}(\alpha^*+\alpha) &
  \frac1\beta G_0 ^{-1}(\omega_{\n};\kb)+\frac{u_0}{2}|\alpha|^2 &
 \frac{\lambda}{2}(\alpha^*+\alpha)  &  \frac{u_0}{4}|\alpha|^2 &
 0 \\
0 & 0 & \frac{u_0}{4}|\alpha|^2 & \frac{\lambda}{2}(\alpha^*+\alpha) &
 \frac1\beta G_0 ^{-1}(\omega_{\n};\kb+\kb_0)+\frac{u_0}{2}|\alpha|^2 &
 \frac{\lambda}{2}(\alpha^*+\alpha)  &  \frac{u_0}{4}|\alpha|^2 \\
\dots & \dots & \dots & \dots & \dots & \dots & \dots 
\end{array} 
\right)
\end{align}
\end{widetext}
which is just the matrix (\ref{matrix_full}) calculated for $a_{\n}=a_{\n}^{\rm MF}$.
Furthermore, the matrix
\begin{align} 
\mathcal{M}_{\alpha}^{\prime}=\mkern150mu\nonumber\\
\left( 
\begin{array}{ccccccc}
\dots & \dots & \dots & \dots & \dots & \dots & \dots  \\
 \frac{u_0}{4}\alpha & \frac{\lambda}{2} &
 \frac{u_0}{2}\alpha &
 \frac{\lambda}{2}  &  \frac{u_0}{4}\alpha &
 0 & 0\\
0 & \frac{u_0}{4}\alpha & \frac{\lambda}{2} &
\frac{u_0}{2}\alpha &
 \frac{\lambda}{2}  &  \frac{u_0}{4}\alpha &
 0 \\
0 & 0 & \frac{u_0}{4}\alpha & \frac{\lambda}{2} &
 \frac{u_0}{2}\alpha &
 \frac{\lambda}{2}  &  \frac{u_0}{4}\alpha \\
\dots & \dots & \dots & \dots & \dots & \dots & \dots 
\end{array} 
\right)
\end{align}
is the derivative of (\ref{matrix_full}) with respect to $a_{\n}^*$,
calculated again for $a_{\n}=a_{\n}^{\rm MF}$.
Note that the MF matrix $\mathcal{M}_\alpha(\omega_{\rm
  n};\kb)$ is diagonal in Matsubara space. This is due to the fact that in MF we
have a stationary classical cavity field which has no temporal fluctuations
which could couple different Matsubara sectors in the atomic field.
This also implies that the last term in the
action (\ref{S_eff}) does not contribute within the MF approach, since the
sum is restricted to $\rm n,m\neq 0$.  

Let us briefly discuss the structure of the coupled MF equations (\ref{mf_eqs}).
The first equation is the MF equation for the condensate spinor and
actually corresponds to an infinite set of equations, one for each Bloch
band (each component of $\Phi_0$). In the numerical solution, one has to introduce a cutoff
in the number of bands in the problem which also sets the size of the
Nambu matrices introduced above. In order to have cutoff-independent
results, the number of bands $\nu$ has to be chosen
such that $\nu E_{\rm R}\gg \kbz T$.
The second equation in (\ref{mf_eqs}) is the MF equation for the
cavity condensate. It contains both the coupling with the condensed
atoms (second term) and to the thermal atoms (last term).

In the MF equations (\ref{mf_eqs}), we assumed the fields $\alpha$ and
each component of $\Phi_0$ to be real. The fact that the phase 
of both the cavity and the condensed atom field is fixed in our 
problem becomes evident by generalizing the real valued order
parameters $\alpha$ and $\Phi_0$ to include an arbitrary phase
$\varphi_\alpha=\arg(\alpha)$ and  $\varphi_{\mathbf{0}}=\arg(\phi_{\mathbf{0}})$. 
This leads to a change of the MF free energy  (\ref{s_mf}) which describes
Josephson like coupling terms which lock the phases of the cavity, the pump laser and 
the BEC together. These phase locking terms results from the two
photon processes involving the pump laser (see lower diagram in
Fig.~\ref{effective_vertices}) and are of two kinds.
The first kind of Josephson term comes from the tracelog part in the free
energy (\ref{s_mf}) and involves thermal atoms only. In order to understand its structure, it is useful
to expand the tracelog in powers of $\alpha$, resulting in
\begin{align}
\label{eq:cav_pump_phaselock}
E_{\rm J}^{\rm th}(\alpha)&=-V\ \frac{\lambda^2}{4}\Pi(0;\kb_0)|\alpha|^2\cos^2\left(\varphi_\alpha-\varphi_{\rm
  p}\right)\nonumber\\&+O\left(|\alpha|^4\cos^4\left(\varphi_\alpha-\varphi_{\rm
  p}\right)\right)\ ,
\end{align}
where we have explicitly reintroduced the phase of the pump
laser $\varphi_{\rm p}$. The latter actually vanishes in the frame rotating
with the pump so that the cavity phase appears alone. % However, one has
% to think of $\varphi_\alpha$ as the phase relative to the pump. 
The function $\Pi(0;\kb_0)$, which is defined in Eq.~(\ref{lindhard}) below
is positive and thus the thermal Josephson
coupling $E_{\rm J}^{\rm th}$, which involves only even powers of $\cos\left(\varphi_\alpha-\varphi_{\rm
  p}\right)$,  is negative definite. As a result, it is minimized whenever
the phase of the cavity relative to the pump is either zero or $\pi$,
corresponding to the atomic momentum distribution being peaked either
at plus or minus $\kb_0$, respectively. 
%The Josephson energy (\ref{eq:cav_pump_phaselock}) vanishes for $T\to 0$ since $\beta\Pi(0;\kb_0)\to 0$.

In the self-ordered regime below the Bose-Einstein condensation temperature, where both $\alpha$ and 
$\Phi_0$ are non-vanishing, a further Josephson coupling arises from the last
term in the free energy (\ref{s_mf}). Introducing in addition the phase $\varphi_{\kb_0}$
of the finite momentum Fourier components of the condensate, it reads
\begin{equation}
\label{eq:cav_pump_BEC_phaselock}
E_{\rm J}^{\rm BEC}(\alpha,\Phi_0)=4N\ \lambda\ |\alpha||\phi_0||\phi_{\kb_0}|\cos\left(\varphi_\alpha-\varphi_{\rm
  p}\right) \cos\left(\varphi_{\kb_0}-\varphi_{\mathbf{0}}\right)\ .
\end{equation} 
For simplicity, we truncated the condensate
spinor only to include $\pm\kb_0$ components and 
also used the fact that the energy is
minimized when the two opposite components of the condensate
spinor are the same: $\phi_{-\kb_0}=\phi_{\kb_0}$.
From Eq.~(\ref{eq:cav_pump_BEC_phaselock}), we see that $E_{\rm J}^{\rm BEC}$ is
minimized when the phase of the cavity field relative to the pump
laser and the one of the $\kb_0$ component of the condensate
relative to the homogenous component are locked such that one of the
two is zero and the other is $\pi$. The Josephson energy
(\ref{eq:cav_pump_BEC_phaselock}) thus implies a phase locking between the
above relative phases as long as we have a BEC.
The $\pi$-locking of the relative phase between the condensate and the
cavity field is in fact analogous to what happens in an interacting BEC,
where the $\kb=0$ condensate and the $(\kb,-\kb)$ pairs of bosons at finite
momentum $\kb$ are locked together by an extensive internal Josephson
coupling due to the anomalous interaction terms $\sim V_k\, \hat{a}_k\hat{a}_{-k} \hat{a}_0^{\dagger}\hat{a}_0^{\dagger}+h.c.$.

Experimentally, the relative phase between the pump and the cavity
field can be observed by recombining the laser beam and the photons
leaking out of the cavity mirrors inside an interferometer. The $0-\pi$ phase
locking between the pump and the cavity originating from $E_{\rm J}^{\rm th}$
has been experimentally observed both in the
thermal gas \cite{vuletic_2003} and in the BEC \cite{eth_2010}
self-organization. By contrast, the additional phase locking
orginating from $E_{\rm J}^{\rm BEC}$ should be present only in the
self ordered regime of the BEC~\cite{eth_2010}, but this has not been investigated so far. 

\vspace{0cm}
\section{Saddle-point solution}
\label{MF}

We now  present our results from a self-consistent numerical solution of 
Eqs.~(\ref{mf_eqs}). We first present the global phase diagram, then discuss the features 
at the onset of spatial order, and finally describe the dynamical band structure across the 
phase diagram.

\subsection{Phase diagram}

\begin{figure}[t]
\vspace{-1.5mm}
\includegraphics[scale=0.39]{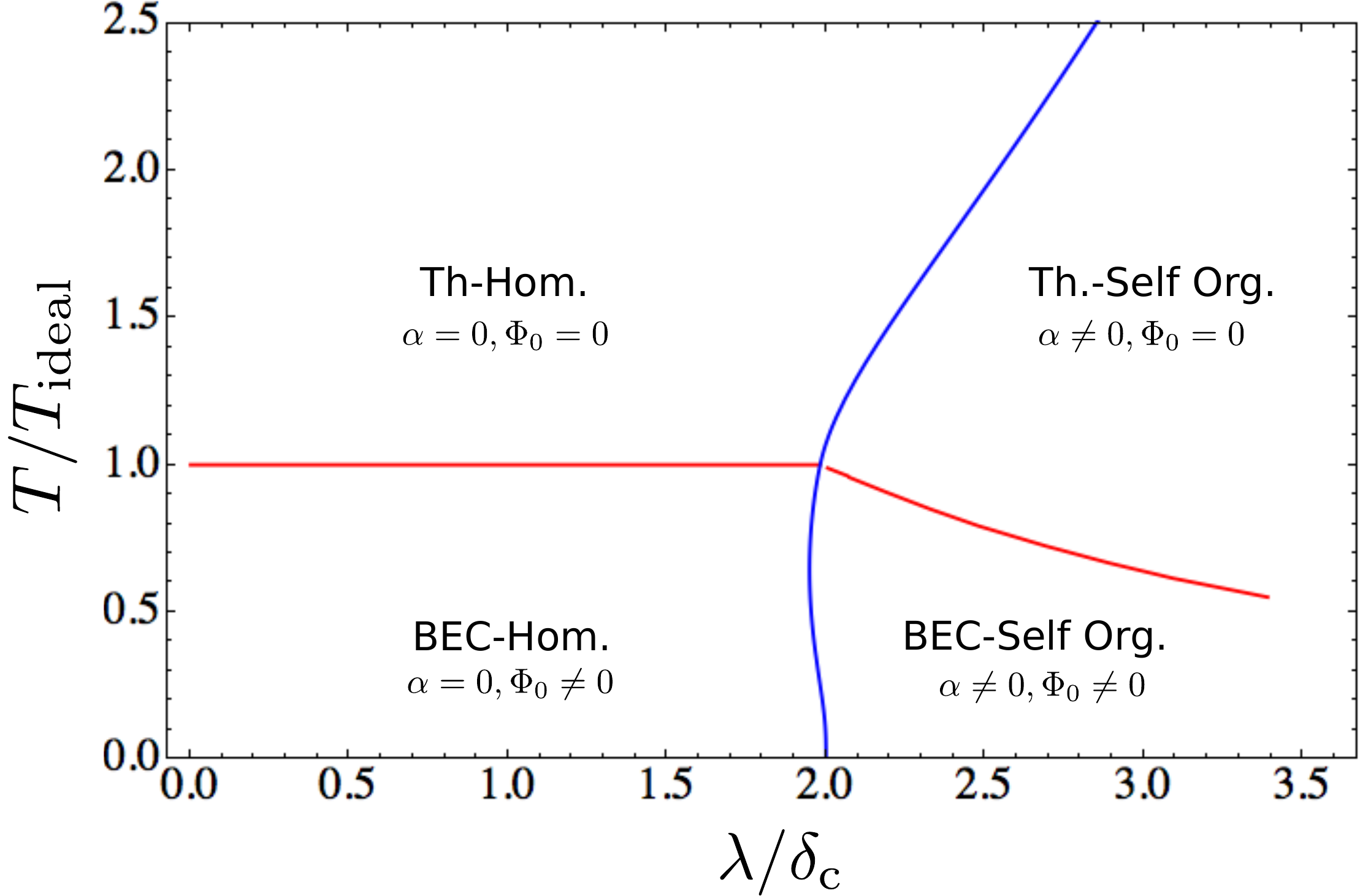}
\caption{Finite temperature ($T$) phase diagram of a Bose gas in an 
optical cavity as a function of atom-cavity coupling $\lambda$. The self-organization
  threshold $\lambda_{\rm so}$ is indicated by the blue line while the ratio
of the condensation temperature $T_0$ to the ideal gas value $T_{\rm
  ideal}$ is indicated by red line. The crossing of these lines happens at the bi-critical point 
  where the atoms become superfluid and spatially self-organize at the same time. 
  Here the recoil energy $E_{\rm R}=8\delta_{\rm c}$, the dispersive shift $u_0=0$,
  and the dimensionless density $\tilde{n}=1$, where
  $\tilde{n}=n/(m\delta_{\rm c})^{3/2}$. }
\label{phase_diag}
 \end{figure}

Upon solving the MF equations (\ref{mf_eqs}),~(\ref{mf_eos}) for different values of
the temperature $T$ and the Rabi coupling $\lambda$ we obtain the
phase diagram shown in Fig.~\ref{phase_diag}. It exhibits four different phases
which are characterized by the two order
parameters $\alpha$ and $\Phi_0$ describing whether and which of the two
symmetries $U(1),Z_2$ (see section \ref{sec:symmetries}) of our system
is broken. The four phases are separated by the self-organization
threshold $\lambda_{\rm so}$ (blue line) and the condensation
temperature $T_0$ (red line). 

The system can thus be found in i) a
thermal homogeneous phase: $\alpha=0,\Phi_0=0$, ii) a BEC homogeneous
phase: $\alpha=0,\Phi_0\neq 0$ where only the continuous $U(1)$
symmetry is broken, iii) a thermal self-ordered phase:
$\alpha\neq 0,\Phi_0=0$ where only the discrete $Z_2$ symmetry is
broken, or iv) a self-organized BEC phase $\alpha\neq
0,\Phi_0\neq 0$ where both symetries are broken.

We point out that at $T=0$ the equations (\ref{mf_eqs}) become equivalent
to the MF equations derived with a multimode approach in
\cite{domokos_2011}.
In particular, at $T=0$ we recover the value (\ref{threshold_DM}) of the self-organization
threshold, as will be discussed in section \ref{so_th}. 
In contrast to \cite{domokos_2011}, our approach also covers the whole
temperature range from the $T=0$ BEC through the condensation
temperature up to the Boltzmann gas. 

% This low temperature regime is the relevant one for the experiment of
% \cite{eth_2010}. On the other hand, our approach allows to
% describe self-organization also in the opposite regime where the gas is
% thermal like in the experiment of \cite{vuletic_2003}, and also to
% interpolate between these two regimes, where new interesting effects appear.

In the following, we will consider in detail the main features of the
phases introduced above. In
section \ref{so_th} we will study the self-organization threshold
$\lambda_{\rm so}$. In section \ref{dyn_band} we will analyze the
dynamical band structure appearing in the self-organized phases, giving rise to
a nontrivial behavior of the condensation temperature $T_0$.

\subsection{Self-organization threshold}
\label{so_th}

In the phase diagram of Fig.~\ref{phase_diag}, we observe an
interesting non-monotonic behavior of the self-organization threshold
$\lambda_{\rm so}(T)$ depicted as the blue line. Indeed, as the
temperature increases from zero, $\lambda_{\rm so}$ decreases over a
certain range of $T$, that is, the self-organization is favored by
thermal fluctuations. This effect, which can be much more pronounced
than in the example of Fig.~\ref{phase_diag}, originates from the thermal occupation
of the momentum continuum and will be now discussed in detail. 
\begin{figure}[]
\vspace{0.0cm}
\includegraphics[scale=0.41]{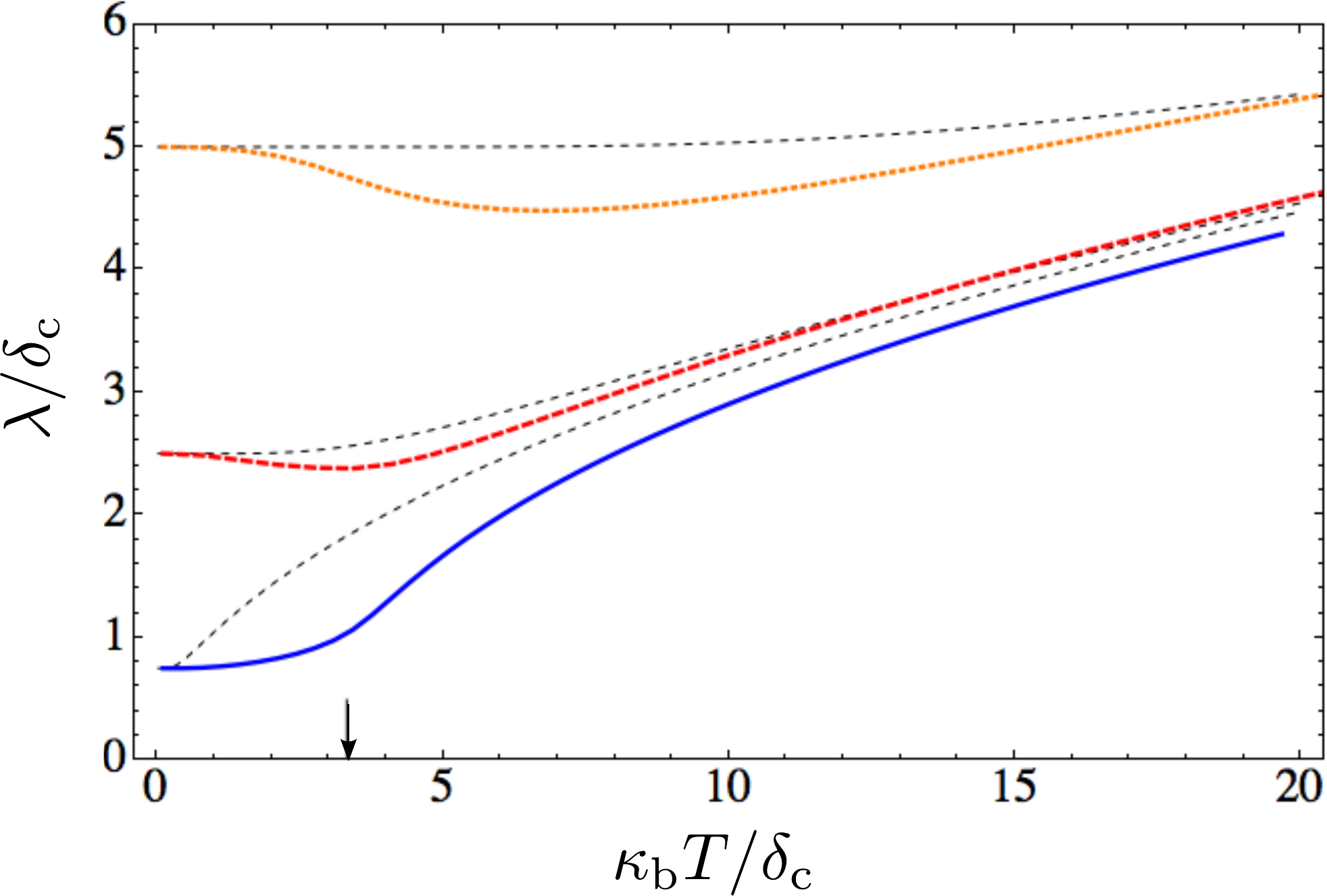}
\caption{The self-organization threshold $\lambda_{\rm so}$ as a function of temperature
  for three different values of the recoil energy:
  $E_{\rm R}=1.1\delta_{\rm c}$ (blue solid line),
  $E_{\rm R}=12.5\delta_{\rm c}$ (red dashed line), and
  $E_{\rm R}=50\delta_{\rm c}$ (orange dotted line). The other
  parameters are $\tilde{n}=1$ and $u_0=0$. The black dashed lines show the
  correspondent prediction from the effective Dicke model
  $\lambda_{\rm D}$. The arrow marks the value of the condensation
  temperature $T_0$.}
\label{threshold_plot}
\end{figure}

It is possible to derive an analytical expression for
$\lambda_{\rm so}(T)$ by linearizing the MF equations (\ref{mf_eqs}) with
respect to $\alpha\ll 1$. Correspondingly, one has to truncate the condensate spinor to
include only the first non-zero momentum component:
$\Phi_0^T=(\dots,0,\phi(-\kb_0),\phi(\mathbf{0}),\phi(\kb_0),0,\dots)$
and linearize the MF equations also with respect to $\phi(\pm\kb_0)\ll\phi(\mathbf{0})$. 
By requiring the coefficient of the linear term in $\alpha$ in the
second equation of (\ref{mf_eqs}) to vanish, we obtain the
self-organization threshold, whose critical rescaled Rabi coupling reads
\begin{equation}
\label{threshold}
\lambda_{\rm so}^2=\frac{\delta_{\rm
    c}}{2(n_0/n)/E_{\rm R}+\Pi(0;\kb_0)/n}\ ,
\end{equation}
with the bosonic Lindhard function
\begin{align}
\label{lindhard}
\Pi(\omega_{\n};\kb)&=\frac{\beta}{V}\sum_{\rm
  n_1}\sum_{\kb_1}G_0(\omega_{\n_1};\kb_1)
G_0(\omega_{\n}+\omega_{\n_1};\kb+\kb_1)\nonumber\\
&=\int_{\mathcal{B}} \frac{d\kb}{(2\pi)^3}
\frac{n_{\rm b}\!\left(\epsilon(\kb)-\mu\right)-n_{\rm b}\!\left(\epsilon(\kb+\kb_0)-\mu\right)}{\epsilon(\kb+\kb_0)-\epsilon(\kb)-i\omega_{\rm
  n}}\ ,
\end{align}
where $n_{\rm b}\!(z)=(\exp(\beta z)-1)^{-1}$ is the Bose distribution function.
%
%We do not present this derivation in more detail since an alternative
%one, based on mode softening in the cavity fluctuation spectrum, will
%be given in section~\ref{self_en}.
The Lindhard function (\ref{lindhard}) for $\kb=\kb_0$ essentially
quantifies how important is the scattering of the cavity photons with
the thermal atoms. As will be discussed in section
\ref{self_en}, it contributes to the cavity photon self-energy. 
The corresponding Feynman diagram is depicted on the second
line (first term) in Fig.~\ref{self_en_all}. In particular, the zero frequency part is the relevant
one for the self-organization threshold (\ref{threshold}). 
On the other hand, the first term in the denominator of (\ref{threshold}) corresponds
to the scattering of the photon with the condensed atoms (see
the second diagram on the second line in Fig. \ref{self_en_all}).
Altogether, self-organization is favored by a more effective scattering between cavity photons and
atoms, that is, if $2(n_0/n)/E_{\rm R}+\Pi(0;\kb_0)/n$ is increased. 

In Fig.~\ref{threshold_plot} $\lambda_{\rm so}(T)$ is plotted for different values of
the recoil energy $E_{\rm R}$, and compared with the prediction
of the effective Dicke model (\ref{threshold_DM}). The main
observation is that, given any recoil energy, the prediction of the effective Dicke
model never reproduces the behavior of $\lambda_{\rm so}$ over
the whole temperature range. As noted in the beginning of
this section, this is due to the two main facts: i) at any finite
temeprature, the momentum distribution of the atoms is a continuum, and ii) the atoms can
Bose condense. The first effect, dominating at large recoil (orange
dotted line) affects $\lambda_{\rm so}$ through the Lindhard
function. The second one, dominating for recoil energies smaller than
the condensation temperature (blue solid line), enters through the temperature dependence of
the condensate occupation $N_0$.
\begin{figure}[]
\vspace{0.0cm}
\includegraphics[scale=0.37]{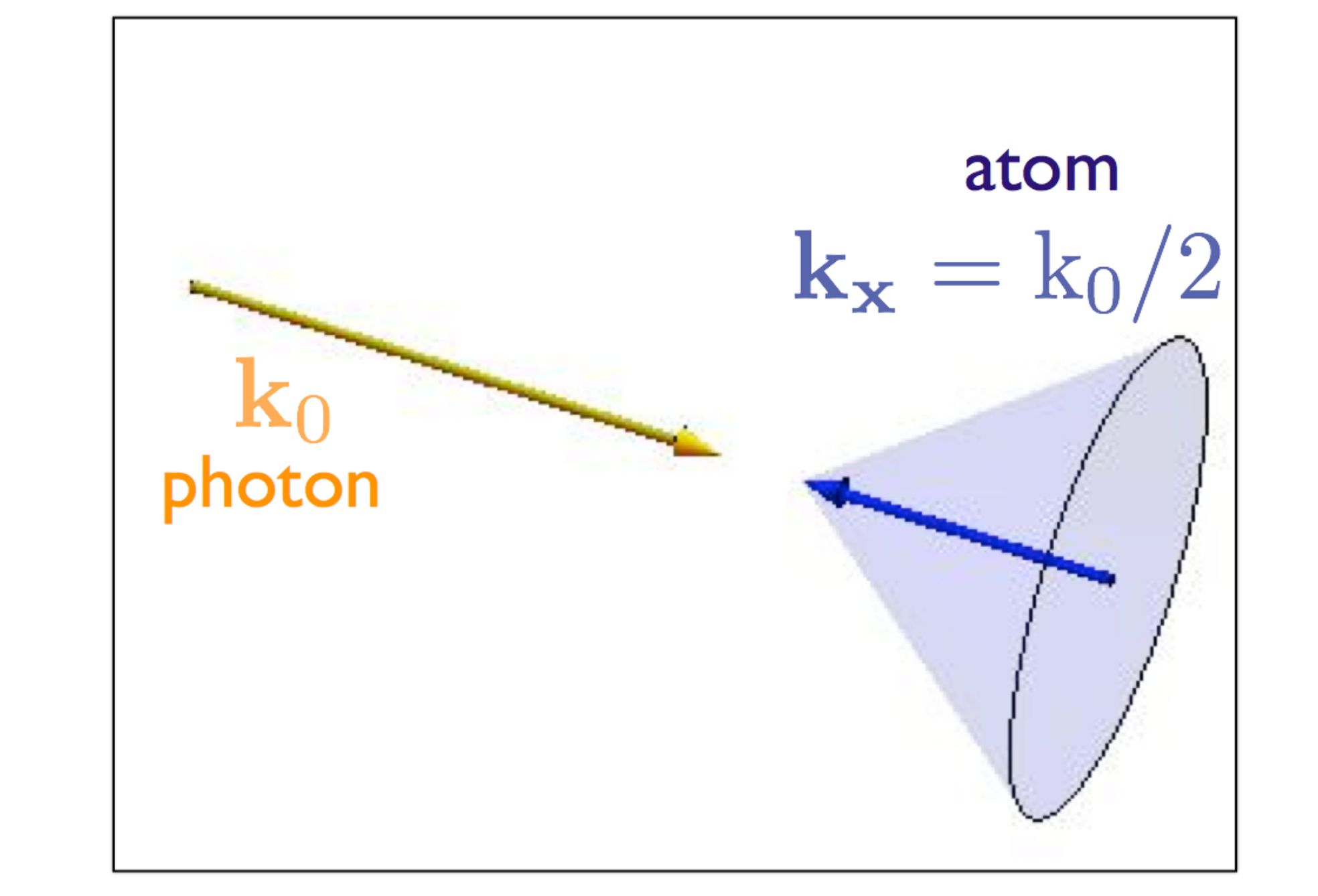}
\caption{Illustration of the momentum configurations of an energetically allowed scattering process between an 
atom and a cavity photon. The arrows
  three-dimensionally represent the
  direction and intensity of the momentum carried by the photon
  (yellow) and the atom (blue). The cone around the blue arrow
  indicates all the other possible directions of the atomic momentum
  giving rise to equivalent contributions to the momentum sum in $\Pi(0;\kb_0)$}
\label{scattering_plot}
\end{figure}

The prediction of the effective Dicke model is recovered only in the
very low or very high temperature regime.
Indeed, if the recoil energy
$E_{\rm R}\gg \kbz T$, we have that $\Pi(0;\kb_0)\simeq
2(n-n_0)/E_{\rm R}$ and $N_0$ disappears from $\lambda_{\rm so}$. The system does not differentiate anymore
between condensed and non-condensed atoms and the rescaled critical
Rabi coupling becomes 
\begin{equation}
\label{th_lt}
\lambda_{\rm so}^2\simeq\frac{\delta_{\rm
    c}E_{\rm R}}{2}\simeq\lambda_{\rm D}^2,\qquad E_{\rm R}\gg \kbz T\ .
\end{equation}
In the opposite high-temperature regime $\kbz T\gg
E_{\rm R},T_0$, we have instead $\Pi(0;\kb_0)\simeq
(n-n_0)/\kbz T$ and $N_0=0$, so that the threshold becomes:
\begin{equation}
\label{th_ht}
\lambda_{\rm so}^2\simeq\delta_{\rm
    c}\ \kbz T\simeq\lambda_{\rm D}^2,\qquad \kbz T\gg E_{\rm
  R},\kbz T_0 \ .
\end{equation}
The high-temperature behavior (\ref{th_ht}) of  the critical coupling agrees precisely
with the result of the semi-classical treatment of Refs. \onlinecite{domokos_2005,domokos_2006}.

We point out that the fact that the self-organization threshold
$\lambda_{\rm so}$ coincides with the Dicke model prediction
(\ref{threshold_DM}) at high temperatures does not mean that the Dicke
model correctly describes the system in this regime. In fact, the atomic 
momentum distribution is a continuum and cannot be truncated to contain 
only $\kb=0$ and $\pm\kb+0$. This does not show up in the result for 
$\lambda_{\rm so}$ because the latter depends only on the
zero-frequency part of the Lindhard function, which for $\kbz T\gg
E_{\rm R}$ does not depend anymore on the recoil energy.
% The ability to tune the detuning $\delta_c$ over a range of magnitudes should make these features accessible in experiments provided one can prepare the atomic gas in the cavity for different 
% temperatures.

In order to experimentally probe both these limiting temperature regimes, and in
particular the intermediate regime where novel features beyond the
Dicke model appear, one would need to prepare the system at different
temperatures both below and above the recoil energy. This should be
possible with cold atoms where typically both the critical temperature
scale $\kbz T_0$ and the recoil energy $E_{\rm R}$ are in the $\mathrm{KHz}$ range.
\begin{figure}[]
%\vspace{0.0cm}
\includegraphics[scale=0.46]{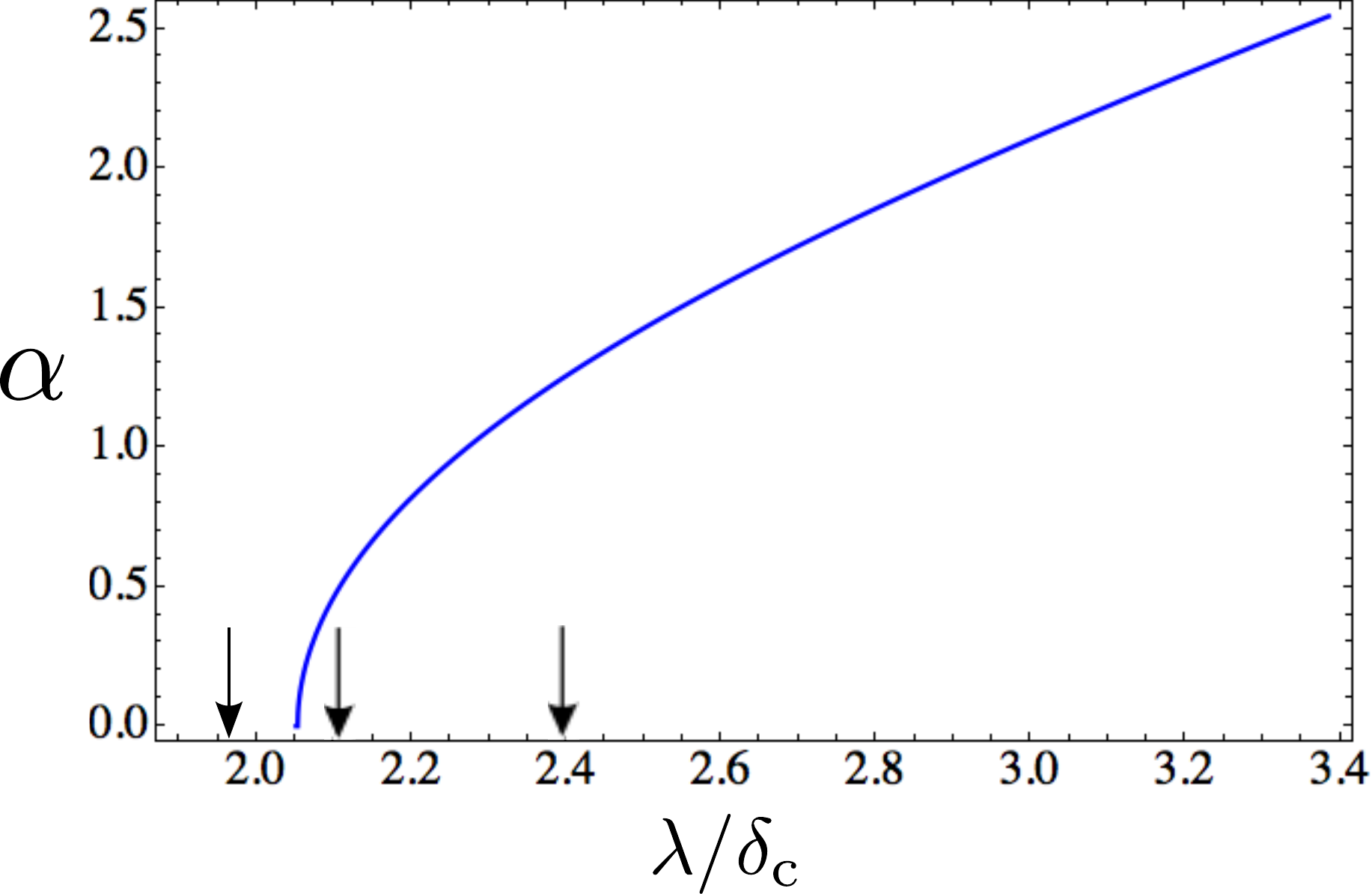}
\caption{Onset of the coherent cavity
  condensate $\alpha$ as a function of the Rabi coupling
  $\lambda$ at a given temperature $\kbz T=4\delta_{\rm c}>T_0$. The parameters
are the same as in Fig.~\ref{phase_diag}.}
\label{alpha_grow}
\end{figure}

\begin{figure*}[]
\vspace{0.0cm}
\includegraphics[scale=0.43]{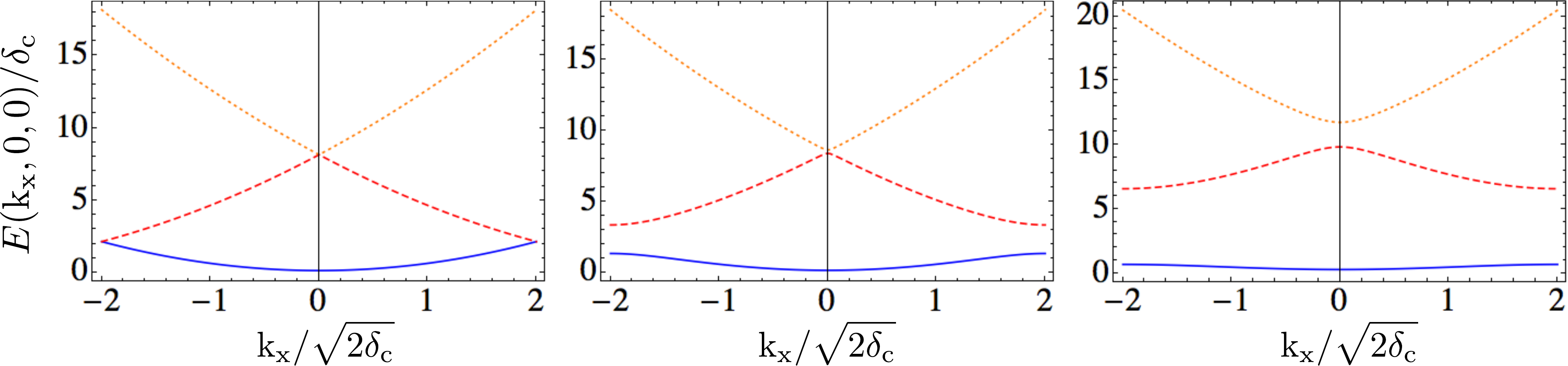}
\caption{Dynamical band structure across the self-organization
  transition (from left to right) displaying the opening of band gaps induced by the coherent cavity field. 
  The three panels show the low band structure for three
  different values of $\lambda$, marked by the arrows in Fig.~\ref{alpha_grow}. The parameters
are the same as in Fig.~\ref{alpha_grow}.}
\label{transition_example}
\end{figure*}

An interesting novel feature that
we observe is the existence of an optimal temperature for which
$\lambda_{\rm so}(T)$ is minimum. This feature, already noticeable in the
phase diagram of Fig. \ref{phase_diag}, becomes more pronounced for
recoil energies large compared to $\kbz T_0$, as apparent from Fig.~\ref{threshold_plot}.
As anticipated in the beginning of this section, the existence of a
minimum for $\lambda_{\rm so}(T)$ is an effect of thermal occupation of
the momentum continuum and is due to the fact that, over a certain
range of temperature, $\Pi(0;\kb_0)$ grows with $T$. 

Let us consider the Lindhard function as written on the second line of
(\ref{lindhard}) and put $\omega_{\rm n}=0$.
For the present argument, the relevant contribution to the momentum
integral is coming from
momenta such that the denominator is
zero. There is no singularity since the numerator is also zero,
yielding a finite value. This corresponds to an event where the atom has the same
energy before and after scattering with the photon (see again section
\ref{self_en}).
This happens every time that the scattering takes place in the
configuration illustrated  in Fig. \ref{scattering_plot}. 
The only condition to be fulfilled in order for the atom not to change
its energy in the scattering is that its momentum component along the
cavity direction is equal to half of the photon momentum:
$\mathrm{k}_x=-k_0/2$.
This means that a whole continuum of momenta satisfying this
condition, indicated by the blue cone in Fig.~\ref{scattering_plot},
give a large contribution to the momentum integral in (\ref{lindhard}).
With increasing the temperature,  this continuum of momenta which differ only in their
component transverse to the cavity axis, acquires a larger occupation with the net effect of increasing
$\Pi(0;\kb_0)$ and thus decreasing $\lambda_{\rm so}$.

However, as can be seen in Fig.~\ref{threshold_plot}, when $T$ becomes
comparable or larger than
$\epsilon(\kb_0/2)=E_{\rm R}/4$, the temperature starts to
disfavor the self-organization. This is due to the occupation of larger
momenta along the cavity direction which subtracts weight from the
$\mathrm{k}_x\sim-k_0/2$ region. 

We thus understand the minimum in $\lambda_{\rm so}(T)$ as resulting
from the competition between the smearing out of the occupation 
of the $k_0/2$ momentum state along the cavity direction and the
occupation of the continuum of momentum states transverse to the
cavity direction.

It is interesting to note that an analogous effect of temperature on
self-organization has been numerically observed also
for a different setup, where the atoms are trapped inside a two-dimensional lattice
and interacting with a cavity mode \cite{hofstetter_2012}.

\subsection{Dynamical band structure}
\label{dyn_band}

An example of the behavior of the cavity condensate $\alpha$ across
the self-organization transition is shown in
Fig.~\ref{alpha_grow}.
In this example the system is above the
condensation temperature $T>T_0$ so that $\Phi_0=0$. 
After the threshold $\lambda_{\rm so}=2.05\delta_{\rm c}$ the cavity
MF $\alpha$ grows like $\sqrt{1-(\lambda/\lambda_{\rm so})^2}$, giving
rise to a lattice felt by
the atoms. This lattice has a wavevector $\kb_0$
with corresponding amplitude $V_{\kb_0}=\lambda\alpha$, and a
wavevector $2\kb_0$ with amplitude $V_{2\kb_0}=u_0\alpha^2/4$. These
amplitudes may be viewed as  ``dynamical'' since they depend on $\alpha$ which in turn
depends on the atomic configuration. 

The corresponding band structure is shown in
Fig.~\ref{transition_example}. We see that, as soon as we cross the
self-organization threshold $\lambda_{\rm so}$, band gaps appear due
to a nonzero lattice depth. The size of the band gaps increases monotonically with the 
distance from the critical coupling. In the example of Fig.~\ref{transition_example}, the
tight binding regime is reached already for $\lambda=1.2\lambda_{\rm so}$ (right panel).

The band structure $E(\kb)$ is obtained by finding the poles of the matrix
$\mathcal{M}_{\alpha}^{-1}(-i\omega;\kb)$ for the real frequency
$\omega$, which amounts to finding the eigenvalues of 
$\mathcal{M}_{\alpha}(0;\kb)$. In general, we have one
pole for each Bloch band. Having found the poles, we
can evaluate the Matsubara sums in the Eqs.~(\ref{mf_eqs}). In
particular, the equation of state becomes:
\begin{equation}
\label{eos_expl}
N=N_0+\sum_{\kb\in\mathcal{B}}\sum_{\nu}n_{\rm b}\!(E_\nu(\kb))\ ,
\end{equation}
where $\nu$ is a positive integer labelling the different Bloch bands.
We note that $N_0=0$ in the example of Fig.~\ref{transition_example}.

The fact that the lattice is dynamical becomes apparent by examining the
equation of state (\ref{eos_expl}) together with the second MF
equation (\ref{mf_eqs}) (assume for simplicty that we have no
condensed atoms $N_0=0$). At given $N,V,T$ and also 
$\lambda,u_0$, the value of $\alpha$ fixes the
band structure $E_\nu(\kb)$ which, through Eq.~(\ref{eos_expl}),
determines how the atoms occupy the different bands. The resulting 
chemical potential enters the second equation in (\ref{mf_eqs}),
determining the amplitude $\alpha$ of the cavity mode. This
feedback loop implies that the band structure here is quite different from
the rigid band structure in optical lattices~\cite{lattice_rmp}: it depends
on how the atoms occupy the bands.

An important consequence of having a dynamical band structure is that the lattice depth at a given $\lambda$ depends on
temperature. The condensation temperature  $T_0$ must be thus 
self-consistently determined and is not directly obtained from
$\lambda$.
An example behavior of $T_0$ is depicted in Fig.~\ref{phase_diag} (red
line). Outside the self-organized phase $\lambda<\lambda_{\rm so}$, the MF
equation of state is the same as the one of an ideal homogeneous Bose gas, and
thus $T_0$ takes the corresponding value $T_{\rm ideal}=3.31(N/V)^{2/3}/m\kbz$ independent of $\lambda$.
In section \ref{spectral_function} we will show that this actually
holds in general (not only in the MF approximation) in the TL.
In the self-organized phase instead, due to the presence of a lattice, $T_0$
starts decreasing with $\lambda$. The depth of the lattice is indeed a
monotonic function of $\lambda$ and a decreasing behavior is thus
qualitatively expected.

\section{Photon dynamics around saddle-point}
\label{spectral_function}

In this section, we compute the cavity spectrum and study its 
temperature dependence across the condensation temperature, as well 
as its behavior close to the DHL self-organization transition.
We will restrict our
analysis to the homogeneous phase $\lambda<\lambda_{\rm
  so}$. In this phase, the cavity coherent field $\alpha=0$ and the atomic condensate is
homogeneous $\Phi_0^T=(\dots,0,\sqrt{N_0/N},0,\dots)$.

\subsection{Photon self-energy}
\label{self_en}

In this subsection, we derive the renormalized cavity photon propagator from which we compute 
the spectral function in the following subsection.

Since $\alpha=0$, we can expand the action (\ref{S_eff}) in the
cavity field $a_{\n}$. Up to second
order, the effective action reads (see appendix
\ref{derivation_propagator} for details of the derivation)
\begin{align}\label{S_eff_2}
S_{\rm eff}^{(2)}[a^*\!,a]&=-\beta\mu
N_0+\sum_{\n}\sum_{\kb}\ln\left(G_0^{-1}(\omega_{\rm
    n};\kb)\right)+\nonumber\\
&\frac{1}{2\beta^2}\sum_{\rm n}
\left(
\begin{array}{cc} a_{\n}^{*} & a_{-\n}^{\phantom{*}}\end{array}
\right)
\mathcal{G}^{-1}(\omega_{\n})
\left(
\begin{array}{c} a_{\n}^{\phantom{*}}\\a_{-\n}^{*}\end{array}
\right)
\ ,
\end{align}
where the inverse cavity propagator reads:
\begin{align}
\mathcal{G}^{-1}(\omega_{\n})/\beta=\mkern180mu\nonumber\\\left(
\begin{array}{cc}
-i\omega_{\n}-\Delta_{\rm c}+\Sigma_{u_0}+\Sigma(\omega_{\n}) & \Sigma(\omega_{\n})  \\ 
\Sigma(\omega_{\n})  & i\omega_{\n}-\Delta_{\rm
  c}+\Sigma_{u_0}+\Sigma(\omega_{\n}) \\ 
\end{array}
\right)\ ,
\label{prop_cav}
\end{align}
with the two kinds of self-energies:
\begin{align}
\Sigma_{u_0}=&\ \frac12 u_0 \label{self_en_disp} \\
\Sigma(\omega_{\n})=&-\lambda^2\left[\frac1{2n}\Pi(\omega_{\n};\kb_0)
  +\frac{(n_0/n) E_{\rm R}}{\omega_{\n}^2+E_{\rm R}^2}\right] \label{self_en_pol}\
.
\end{align}

\begin{figure}[b]
\includegraphics[scale=0.30]{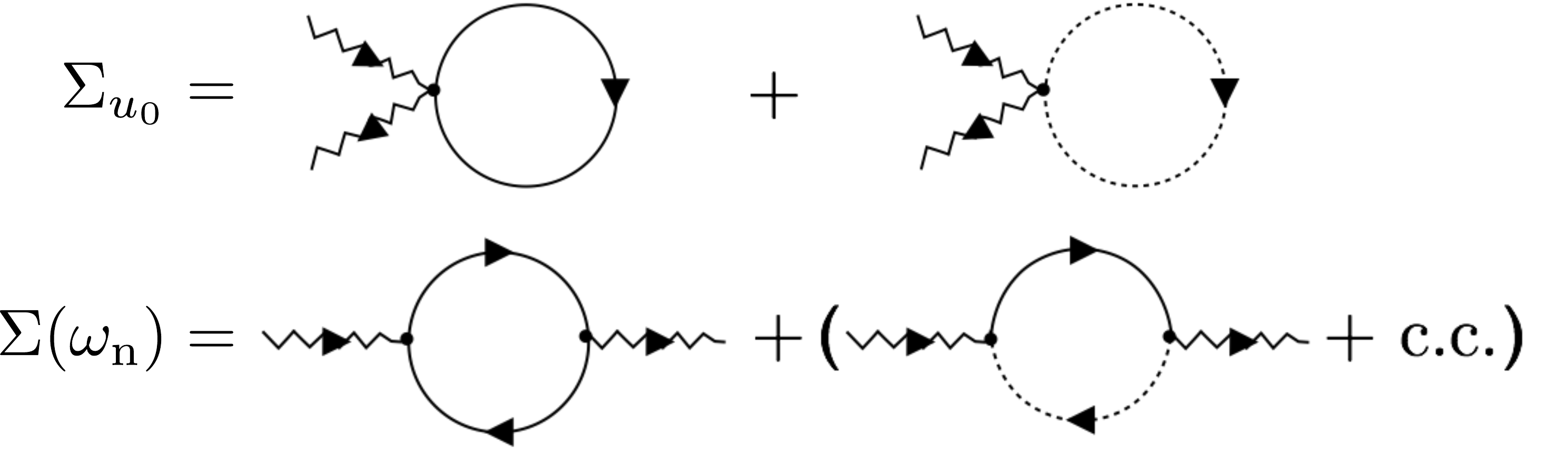}
\caption{Feynman diagrams contributing to the photon self-energy. The
  zig-zag lines are for photons, the straight solid lines for thermal
  atoms, and the dashed lines for condensed atoms. }
\label{self_en_all}
\end{figure}

The first self-energy, whose corresponding Feynman diagrams are shown
on the first line in Fig.~\ref{self_en_all}, results from events without momentum transfer
between the photon and the atoms. These
correspond to Hartree-terms proportional to the average single-atom dispersive
shift $u_0$.
We have one single internal
atomic line proportional to $G_0(\omega_{\rm s};\kb)$ for thermal
atoms or proportional to $N_0$ for condensed atoms. For thermal atoms,
upon performing the internal loop frequency and momentum sum, we obtain simply $n-n_0$ and
thus the two contributions to $\Sigma_{u_0}$ in Fig.~\ref{self_en_all} sum
up to yeld an overall dispersive shift, as given in
Eq.~(\ref{self_en_disp}).
\begin{figure*}[]
%\vspace{0.5cm}
\includegraphics[scale=0.48]{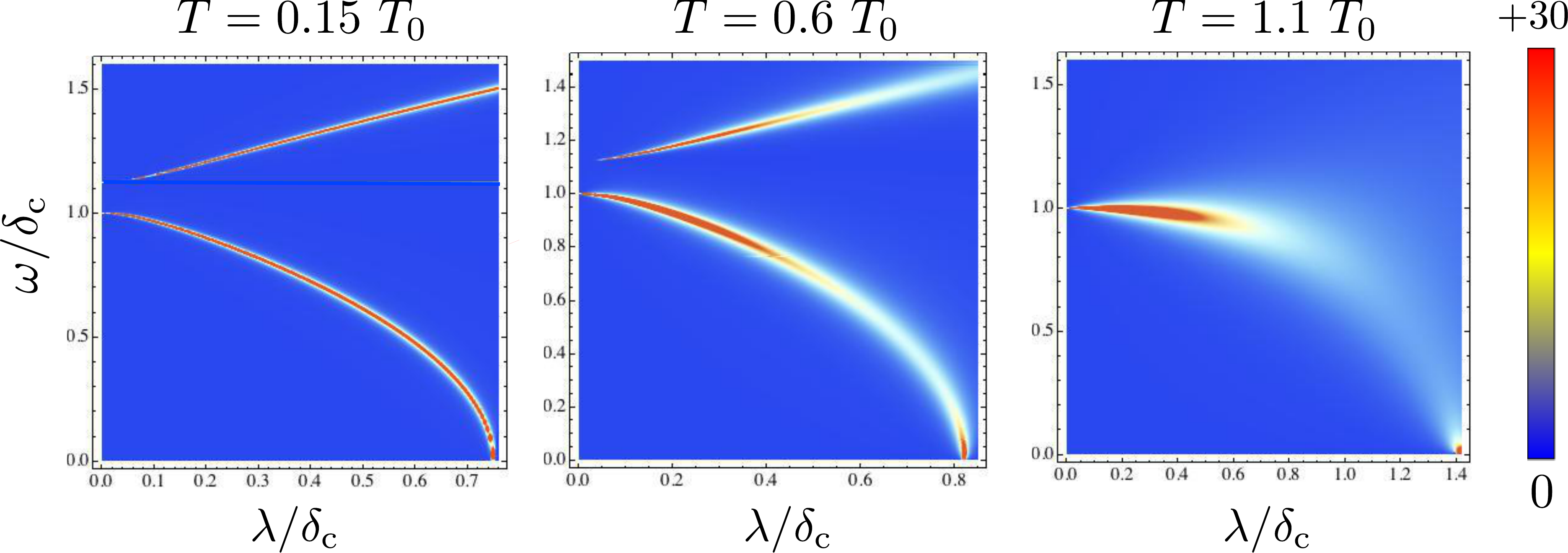}
\caption{Behavior of the photon spectral function with increasing
  temperature. The two-dimensional color plots show $A(\omega)$ as a
  function of the coupling $\lambda$ for different values of the
  temperature. Here the condensation temperature is
  $T_0=3.31\delta_{\rm c}$ and the recoil energy is
  $E_{\rm R}=1.125\delta_{\rm c}$. Also $u_0=0$ and $\tilde{n}=1$.
}
\label{SF_BEC_plot}
\end{figure*}

The second self-energy (\ref{self_en_pol}) involves events
where the photon transfers momentum $\kb_0$ either to a thermal atom
(first term on the second line in Fig.~\ref{self_en_all}) or to a
condensed atom (second term on the second line
in Fig.~\ref{self_en_all}). As anticipated in section \ref{so_th}, the
former diagram corresponds to the Lindhard function defined in
Eq.~(\ref{lindhard}) for $\kb=\kb_0$. Indeed, we have two internal
lines for thermal atoms each proportional to $G_0$ but with frequency
and momentum differing by the photon frequency $\omega_{\n}$ and the
photon momentum $\kb_0$. 
Finally, the second term on the second line in Fig.~\ref{self_en_all} contains a
thermal atomic line $G_0$ with the same frequency and momentum as the
photon, and also a condensed line proportional to $N_0$. This diagram is imaginary
and sums up with its complex conjugate to yeld $N_0$ times the real part of $G_0$,
as given in the second term of Eq.~(\ref{self_en_pol}).

From the propagator (\ref{prop_cav}) we can immediately derive the
expression (\ref{threshold}) for the self-organization threshold
$\lambda_{\rm so}$ by requiring that
$\rm{Det}\!\left[\mathcal{G}^{-1}(0)\right]=0$. 
This amounts to requiring the existence
of a zero frequency mode in the excitation spectrum, i. e. a soft mode. The
properties of the spectrum will be discussed in detail in section
\ref{SF} by means of the spectral function.

Although in the following we will always use the general form of the self-energies
given in Eqs.  (\ref{self_en_disp}) and (\ref{self_en_pol}), it is
interesting to look at their behavior in the limiting regime of large
recoil energy $E_{\rm R}\gg \kbz T$. 
In this case, the Lindhard function simplifies to $\Pi(\omega_{\n};\kb_0)\simeq
2(n-n_0) E_{\rm R}/(\omega_{\n}^2+E_{\rm R}^2)$ and sums
up with the condensed atom contribution in the self-energy
(\ref{self_en_pol}), to yield
$\Sigma(\omega_{\n})\simeq-\lambda^2
  E_{\rm R}/(\omega_{\n}^2+E_{\rm R}^2)$,
which does not depend on the condensate occupation $N_0$ anymore.
As noted also in section \ref{so_th}, this means that
in this regime the system does not differentiate between condensed and
non-condensed atoms and we recover the physics of two-level atoms with
the effective resonance $E_{\rm R}$.

\subsection{Cavity spectral function}
\label{SF}

\begin{figure}[b]
%\vspace{0.5cm}
\includegraphics[scale=0.4]{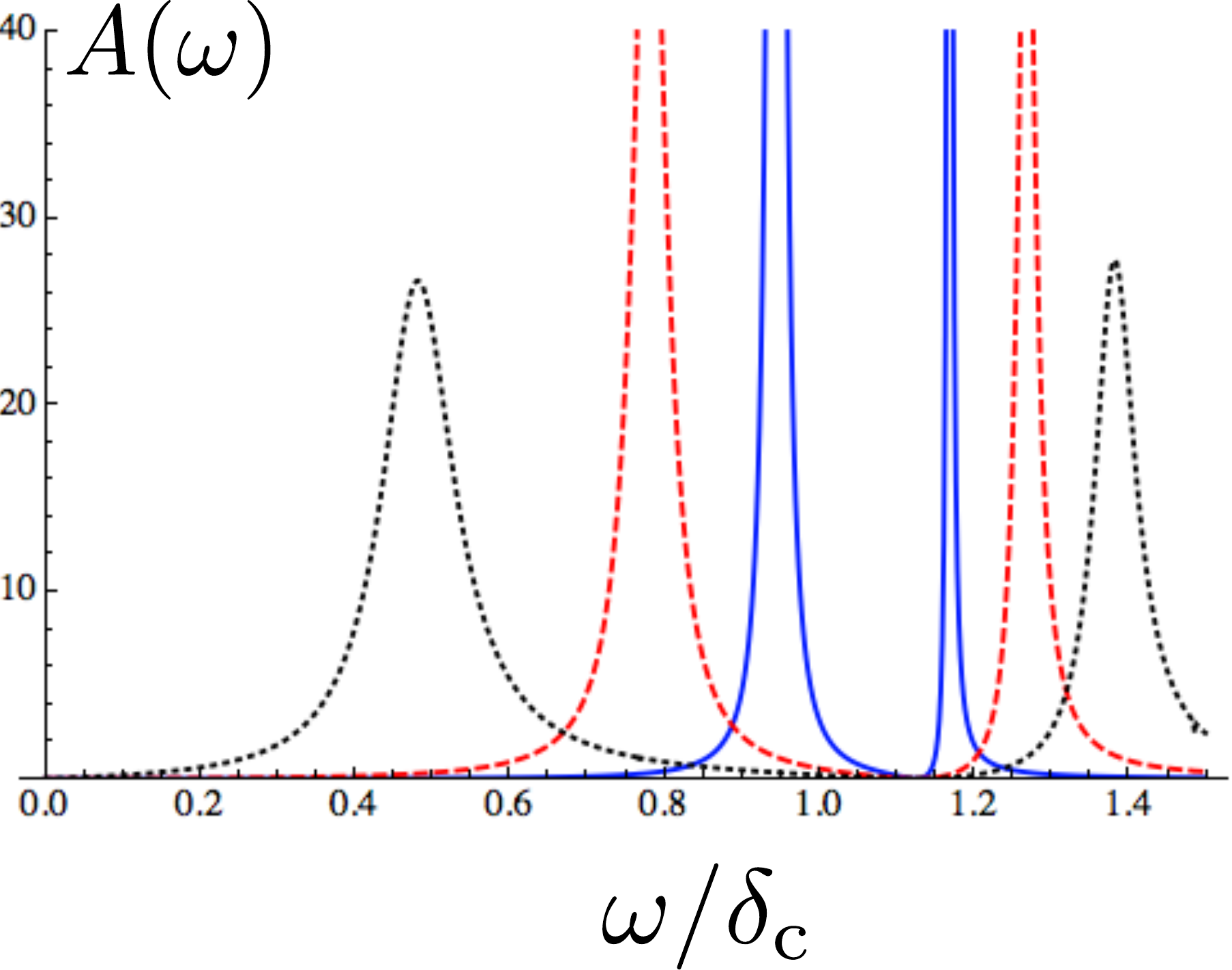}
\caption{The spectral function for
  the case $T=0.6T_0$ at fixed $\lambda=0.17\delta_{\rm c},
  0.41\delta_{\rm c}, 0.67\delta_{\rm c}$ for the blue-solid,
  red-dashed, and black-dotted line, respectively. 
}
\label{SF_BEC_plot_cut}
\end{figure}
We now want to study the spectral function of our effective cavity
below the self-organization transition. 
From the cavity
propagator defined in Eq. (\ref{prop_cav}), we obtain the spectral function
\begin{equation}
\label{SF_def}
\mathcal{A}(\omega)=2\mathrm{Im}\mathcal{G}(-i\omega+0^+)
\end{equation}
by analytic continuation. We will here discuss the diagonal element of 
the two by two matrix $\mathcal{A}(\omega)$, namely: 
\begin{align}
A(\omega)=\mkern180mu\nonumber\\\frac{-2\left(\delta_{\rm
      c}+\omega\right)^2\mathrm{Im}\Sigma(-i\omega+0^+)}{\left[\delta_{\rm
    c}^2-\omega^2+2\delta_{\rm
    c}\mathrm{Re}\Sigma(-i\omega+0^+)\right]^2+\left[2\delta_{\rm
    c}\mathrm{Im}\Sigma(-i\omega+0^+)\right]^2}\ ,
\end{align}
which corresponds to the analytical continuation of the normal propagator
$G_{11}(\tau)=\langle T\hat{a}(\tau)\hat{a}^\dag\rangle$, where
$T$ is the time ordering operator.
The spectral function $A(\omega)$ describes how
the spectral weight of the bare cavity mode is split and broadened by
the coupling with the atoms through the driving laser field. As
done in the experiment \cite{eth_soft}, this can be probed by
exciting the cavity with a weak pulse.
\begin{figure*}[]
\vspace{0.5cm}
\includegraphics[scale=0.57]{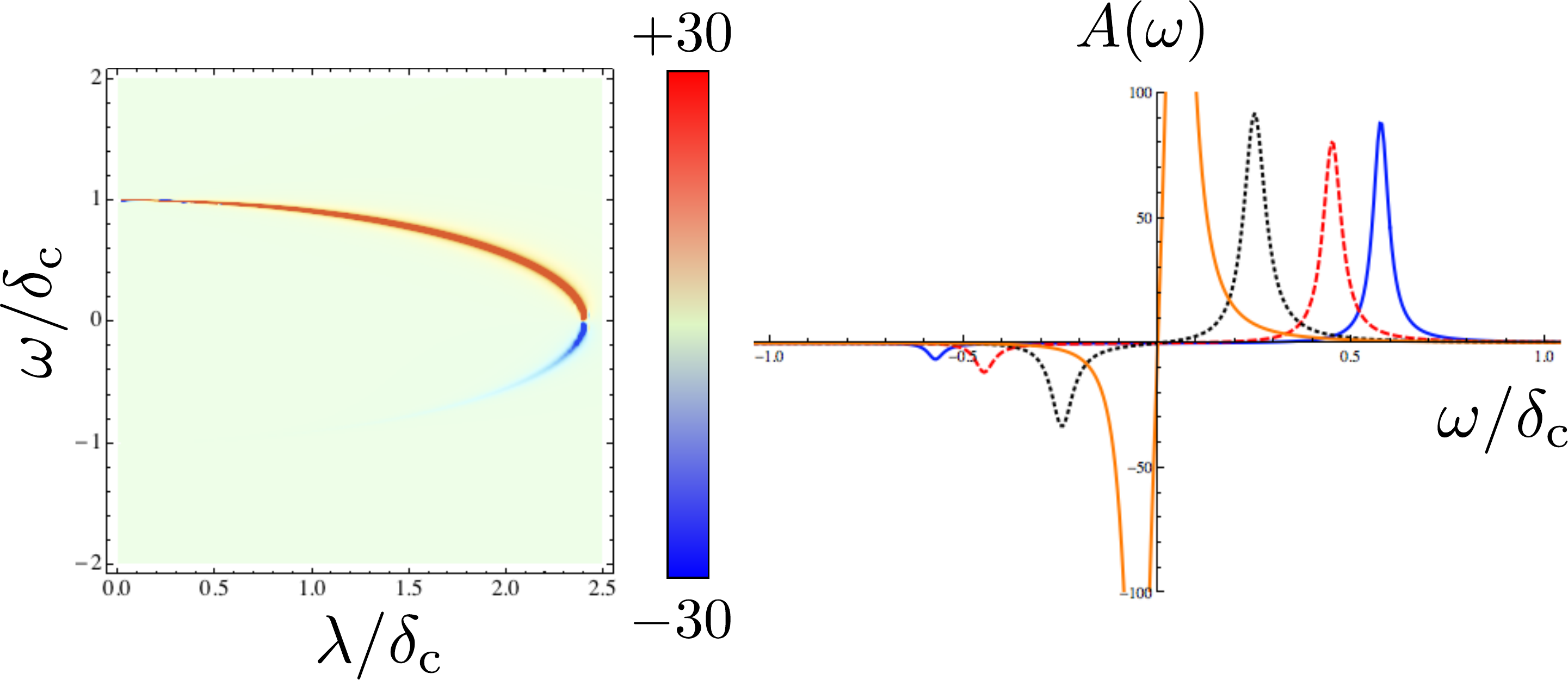}
\caption{Behavior of the spectral function towards the
  self-organization transition. The two-dimensional colour plot shows $A(\omega)$ as a
  function of the coupling $\lambda$. The right panel shows the spectral function for
  at fixed $\lambda=1.97\delta_{\rm c},
  2.15\delta_{\rm c}, 2.33\delta_{\rm c}, 2.40\delta_{\rm c}$ for the blue-solid,
  red-dashed, black-dotted line, and orange solid line respectively. Here the the recoil energy is
  $E_{\rm R}=12.5\delta_{\rm c}$ and $T=4\delta_{\rm c}$. Also $u_0=0$ and $\tilde{n}=1$.
}
\label{SF_neg_plot}
\end{figure*}

Typical examples of $A(\omega)$ as a function of the coupling
$\lambda$ are given in Fig.~\ref{SF_BEC_plot} and \ref{SF_BEC_plot_cut}. 
In general, we observe two peaks corresponding to the two polaritonic branches: the atomic
branch, starting from the value $\omega=E_{\rm R}$ for $\lambda=0$ with
zero spectral weight, and the photonic branch, starting as a
delta-peak at the value $\delta_{\rm c}$ for $\lambda=0$.
The peak corresponding to the lowest branch (the photonic one in
the figures) is then shifted towards $\omega=0$ as lambda
increases and reaches zero at $\lambda_{\rm so}$. This mode softening,
which has been demonstrated experimentally \cite{eth_soft}, has some interesting features that we will discuss at the end of this
section. 

Let us start by considering the effect of temperature plus the momentum
continuum on the spectral
function. In all the examples presented in Fig.~\ref{SF_BEC_plot} we
have a broadening of the polaritonic peaks, which is also evident by
looking at the spectral function at given $\lambda$, as shown in Fig.~\ref{SF_BEC_plot_cut}. 
The broadening is induced by
the coupling of the cavity with
thermal atoms such that a polariton can decay by exchanging momentum
and energy with the latter. 
Formally, this is due to the fact that
the analytic continuation yelds an imaginary part to the
self-energy $\Sigma(-i\omega+0^+)$ through the Lindhard function
(\ref{lindhard}). Indeed we have:
\begin{equation}
\label{lindhard_im}
\mathrm{Im}\Pi(-i\omega+0^+;\kb_0)=\frac{\beta\pi^3}{\ell_T^4k_0}\ln\left[\frac{e^{\beta\left(\frac{|\omega-E_{\rm R}|^2}{4
          E_{\rm R}}-\mu\right)}-e^{-\beta\omega}}{e^{\beta\left(\frac{|\omega-E_{\rm R}|^2}{4
          E_{\rm R}}-\mu\right)}-1}\right]\ ,
\end{equation}
where $\ell_T=2\pi/\sqrt{2m\kbz T}$ is the thermal wavelength. From eq.~(\ref{lindhard_im}), we see
that the imaginary part vanishes for $T=0$. In general, for
$k_0\ell_T\gg 1$, the broadening is strongly suppressed, as can be
seen in the left panel of Fig.~\ref{SF_BEC_plot}. This is
consistent with the previously observed fact that in this regime we
recover the physics of two-level atoms for an effective Dicke model,
which indeed should have no broadening of the polaritonic peaks in
absence of dissipation.
Upon increasing the temperature the broadening increases especially on
the upper polaritonic branch. In particular, by comparing the
$T=0.6T_0$ with the $T=1.1T_0$ case in Fig.~\ref{SF_BEC_plot}, we can
observe the effect of crossing the condensation temperature $T_0$. The
upper polariton peak gets quickly washed out and a strongly broadened
lower polariton peak remains. This is due to the vanishing of the
condensate $N_0$. The latter contributes to the spectral function with a sharp feature since the
corresponding self-energy (second term in Eq.~(\ref{self_en_pol})),
has no imaginary part. 
However, the upper polaritonic branch is not strictly a signature of
the presence of the atomic condensate. This is true only when the
polariton energy is much smaller than $\kbz T_0$, as in the example of
Fig.~\ref{SF_BEC_plot}. Indeed, in this regime the broadening due to
the coupling with thermal atoms close
to $T_0$ is so large that, without a condensate, the upper
branch will be completely washed away.
On the other hand, if we choose $\delta_{\rm c}$ and the recoil
$E_{\rm R}$ much larger than the temperature, the latter has no
such effect on the polaritons, as long as we stay far enough from the transition. For instance, even at large
temperatures, we recover the usual collective Rabi splitting, whereby
the two equally broadened polaritonic peaks move apart for increasing
(and small enough) coupling.

In Fig.~\ref{SF_BEC_plot} and \ref{SF_BEC_plot_cut}, we see that the
broadening increases with the coupling $\lambda$. However, close
enough to the self-organization threshold, the soft mode peak seems to
become well defined again. This feature is mostly evident in the rightmost
panel of Fig.~\ref{SF_BEC_plot}, where the soft mode peak first is washed out but then
reappears very close to the transition. As we shall discuss now, the
peak is not actually becoming sharper, but acquiring a diverging weight.

Let us analyze the behavior of the spectral function close to the
transition in more detail with the help of
Fig.~\ref{SF_neg_plot}. Here the negative frequency part
of $A(\omega)$ is also shown. Close enough to the
transition, a well defined peak with negative spectral weight appears at
negative frequency. This negative peak compensates the weight coming
from the positive peak thereby enforcing the sum rule $\int_{-\infty}^{+\infty}d\omega A(\omega)=2\pi$. 
This negative frequency branch corresponds to excitations where the cavity 
extracts photons from the pump laser through the interaction via the atoms. 
Indeed, the coupling with the laser gives rise to terms proportional to $a+a^*$ in the
effective action (\ref{S_prime}) after integrating out the excited
atomic state. This terms in turn produce anomalous components in
the propagator (\ref{prop_cav}) even in absence of the cavity
condensate, yielding negative frequency modes with negative spectral
weight.

The closer we get to the transition, the stronger becomes the negative
peak which also moves towards zero frequency in a symmetric way with
respect to the positive frequency part. Close to
$\lambda_{\rm so}$ and for small $\omega$, we can derive an analytical expression for the
spectral function:
\begin{equation}
\label{SF_div}
A(\omega)\simeq\frac{C(T)\omega}{\left[\delta_{\rm
    c}^2(1-(\frac{\lambda}{\lambda_{\rm so}})^2)-2\left(1-(\frac{\lambda}{\lambda_{\rm so}})^2\right)\omega^2\right]+\left(C(T)\omega\right)^2}\ ,
\end{equation}
where $C(T)=\beta^2\lambda_{\rm so}^2 (\pi^3/\ell_T^4nk_0)n_{\rm b}\!(\epsilon(\kb_0/2)-\mu)$.
The above expression illustrates the $1/\omega$ behavior for
$\lambda\to\lambda_{\rm so}$, noticeable in the
right panel of Fig.~\ref{SF_neg_plot}. This can be physically interpreted as a situation in which the cavity can extract photons
from the pump laser with perfect efficiency, which is indeed what
should happen at the self-organization transition.

In the denominator of the spectral function
\eqref{SF_def}, the first term between square brackets determines the
position of the peak in the spectral function, while the
last term sets the width of the peak according to the broadening
factor $C(T)$. However, as we see from
Eq.~\eqref{SF_div}, this interpretation is not valid for the soft mode at
the transition, since the term setting the width has the same
$\omega$-dependence as the one responsible for the peak position. This
is due to the fact that the peak broadening, proportional to the
imaginary part of the self energy \eqref{lindhard_im}, behaves like
$\omega$ at small frequency. This implies that the polaritons
are not well defined quasiparticles for $\omega\to 0$. 
Further insight into this aspect is provided by analyzing the behavior
of the poles of the propagator $\mathcal{G}(-i\omega+0^+)$. These are
determined by the equation: $\delta_{\rm c}^2-\omega^2+2\delta_{\rm
  c}\Sigma=0$. The result for the soft mode pole is:
\begin{equation}
\omega_{-}\simeq\frac{\delta_{\rm
    c}}{2}\left(\sqrt{1-(\frac{\lambda}{\lambda_{\rm
      so}})^2 -C^2(T)}-iC(T)\right)\ .
\end{equation}
Far enough from the transition, the term under the square root is
positive therefore the pole has both an imaginary part, related to the
broadening factor $C(T)$, and a real part. However, close enough to
the transition, the term in the square root must become negative,
thereby making the soft mode pole purely imaginary
\begin{equation}
\omega_{-}\simeq \frac{-i}{C(T)}\frac{\lambda_{\rm
    so}-\lambda}{\lambda_{\rm so}}\ .
\end{equation}
This means that the soft mode becomes overdamped close to the transition with 
the position of the pole vanishing linearly with $\lambda_{\rm so}-\lambda$ along 
the imaginary axis.
This behavior is the same, but the origin of the effect different, as observed in \cite{dallatorre_2012} in a calculation of the 
two-state Dicke model with photon decay. In that reference, the broadening and 
overdamping of the polariton excitations is due to coupling of photons to the bath 
of electromagnetic modes outside the cavity. In the present paper, the polaritons become 
dissipative from coupling to the thermal bath of bosonic atoms. 

The behavior \eqref{SF_div} of the spectral function close to the transition
determines the photon flux exponent
\cite{domokos_2008,oztop12,dallatorre_2012} characterizing the
divergence of the average photon density at the critical
point. Indeed, at thermal equilibrium the following relation holds \cite{altland_simons}:
$\langle\hat{a}^{\dag}\hat{a}\rangle=G_{11}(0^{-})=\int \frac{d\omega}{2\pi}
A(\omega)n_{\rm b}\!(\omega)$. Since the divergence comes from the
small frequency part of the integral, we can approximate
$n_{\rm b}\!(\omega)\simeq 1/\beta\omega$ and perform the integral in an
interval around $\omega=0$ using the expression \eqref{SF_div}, to get:
\begin{equation}
\label{photon_flux}
\langle\hat{a}^{\dag}\hat{a}\rangle\propto\left(\lambda-\lambda_{\rm
    so}\right)^{-1}\ ,
\end{equation}
which gives a photon flux exponent equal to one, typical of a thermal
behavior. The same exponent was also found in \cite{dallatorre_2012},
where the thermal character of the steady state was again to ``thermal'' noise 
from cavity decay. The result \eqref{photon_flux} is valid at
any finite temperature. At $T=0$ instead, since
$A(\omega)\propto\delta(\omega-\omega_{-})$, we have
$\langle\hat{a}^{\dag}\hat{a}\rangle\propto 1/\omega_{-}\propto(\lambda-\lambda_{\rm
    so})^{-1/2}$ as expected \cite{oztop12,nagy11,dallatorre_2012}.

% \section{Exactness of solution in thermodynamic limit}
\section{Mean-field nature of the self-organization transition}
\label{TL}

In this section, we study the scaling of the fluctuations in the 
TL  and show that the
self-organization transition is of the MF type.
Indeed, as we will show below, the corrections to the MF
solution in the self-organized phase vanish in the TL, and in the
homogeneous phase the Gaussian theory for the fluctuations about
the cavity vacuum, corresponding to the action (\ref{S_eff_2}),
becomes exact.
Moreover, we will show that there is no quantum depletion of the
atomic condensate in absence of additional short-range interactions 
between the atoms.

Up to now, we have truncated our expansion of the action (\ref{S_eff}) around the cavity MF at the
second order in the cavity fluctuations to obtain the action
(\ref{S_eff_2}). However, higher order terms, corresponding to
interactions between the cavity photons mediated by the atoms, can in principle be
relevant. For instance, in the framework of an effective
Ginzburg-Landau theory for the cavity field, the higher order terms
are needed to determine the nature of our self-ordering transition.
In this section, we will study the full effective action (\ref{S_eff})
in the TL, where some exact statements can be made.

\begin{figure}[t]
\includegraphics[scale=0.3]{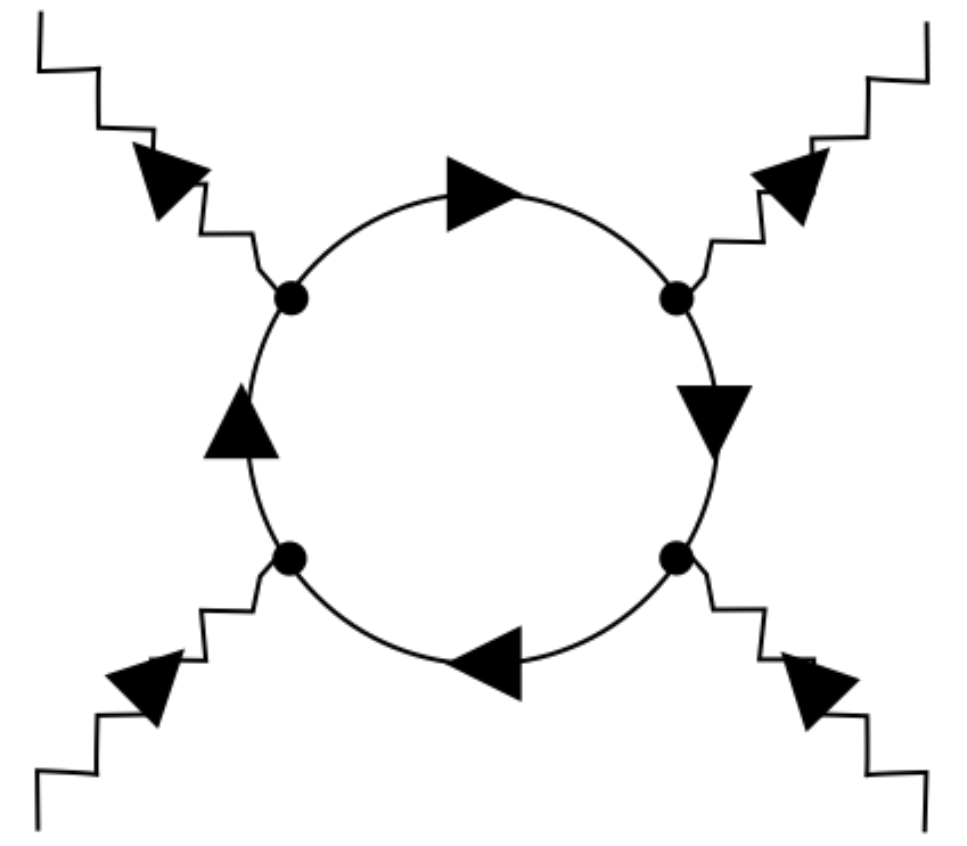}
\caption{Typical $4$-th order Feynman diagram describing the
atom-mediated interaction between cavity photons. The notation is the
same as in Fig.~\ref{self_en_all}.}
\label{diag_interaction}
\end{figure}

% In the TL, the volume of the cavity is scaled together with the volume
% $V$ where the atoms are confined.
% Correspondingly, the bare dipole coupling $g_0$ scales like
% $V^{-1/2}$, thus $\lambda\propto V^{-1/2}$ and $U_0\propto V^{-1}$.
Let us consider the scaling in the TL of the diagrams appearing in our
theory. For instance, consider a typical self-energy diagram, like the one in
Fig.~\ref{self_en_all}. Since for each photonic leg we have a vertex
scaling like $1/\sqrt{N}$, such a diagram scales like unity for the fluctuations
$\omega_{\n}\neq 0$ and like $N$ for the MF (when the external
photonic legs are proportional to $\sqrt{N}\alpha$). Indeed, there is no external loop thus
no scaling from the external legs, apart from the MF diagrams. The same reasoning can be applied to a
typical interaction ($4$-th order) term, like the one in
Fig.~\ref{diag_interaction}. In this case, the addition of two
external legs brings a further $1/N$ scaling due to the vertices, but
no further scaling due to the external lines, apart again from the MF
diagrams. Therefore, a typical $4$-th order diagram scales like $1/N$
for the fluctuations and like $N$ for the MF.

These scaling arguments have several important implications:
First, the MF diagrams contribute to all orders
in the cavity field, i.e. the action (\ref{s_mf}) cannot be truncated
at some power of $\alpha$. Moreover, the fluctuation diagrams, at
most of order one, are negligible with respect to the MF diagrams $\propto
V$. Therefore, the MF approach discussed in section \ref{MF} becomes
exact in the TL. 
This is the same as it happens for the restricted BCS model \cite{muehlschlegel_1962,dukelsky_2004} where the
coupling between fermions takes place only at fixed exchanged
momentum equal to zero. We will come again to this analogy at the end
of this section.
In all our calculations, we found that our MF approach
predicts a continuous self-organization transition.

Clearly, outside the self-organized phase the MF $\alpha=0$ and we are left
with the fluctuation diagrams only. Yet, we are allowed to retain only
the self-energy diagrams of Fig.~\ref{self_en_all} and neglect all the
interaction diagrams which scale like $1/N$. This means that the
action (\ref{S_eff_2}) becomes exact in the TL and we have a
free Gaussian theory. This in turn implies that the cavity propagator
(\ref{prop_cav}) becomes exact in the TL.

A further relevant implication of the above scaling arguments is that
the fluctuations originating from the cavity do not deplete the atomic
condensate in the TL. This is due again to the fact the these
fluctuations are of order unity and are thus negligible with
respect to the number of atoms $N$. In particular, since the cavity
fluctuations are the only quantum fluctuations (surviving even at
$T=0$), we conclude that there is no quantum depletion of the atomic
condensate in the TL. 

We point out that the scaling arguments also apply if one includes
short-range interactions between the atoms. More precisely, the fact
that the mean-field is exact in the TL still holds, even though the model is
not exactly solvable anymore.
The underlying reason for these scalings is the fact that the cavity photons 
have only one fixed momentum. That is, we have one single degree of
freedom $a(\tau)$ in zero dimensions. In this sense, the
self-organization transition is to be regarded as a quantum
bifurcation rather than a quantum phase transition. Indeed, there is
no divergent length scale and no quantum critical behavior, like it
happens in the standard Dicke model.

The exactness of MF theory for the present problem can be 
understood from a different point of view by considering 
the effective interaction between atoms which is mediated by the cavity photons. 
This interaction is obtained from the full action (\ref{S_prime}) by integrating out 
the cavity field $a_{\n}$. Since the latter only appears 
quadratically, this is a Gaussian integration which can formally be done
at the expense of introducing terms beyond quadratic order in the ground state atomic field $\psi_{\n,\kb}$. 
Expanding the resulting action up to
fourth order in $\psi_{\n,\kb}$, one obtains a standard density-density interaction
between the atoms of the form
\begin{align}
\label{long_range_int}
&S_{\rm int}=-\frac{\lambda^2}{2N}\times\nonumber\\&\int_{-\beta/2}^{\beta/2}\!\!\!\!\!\!\!\! d\tau d\tau^\prime \!\!\!\!\int_{V}\!\!\!\! d\rb
d\rb^\prime n(\tau,\rb)\cos(\kb_0\cdot\rb)\mathcal{G}(\tau-\tau^\prime )
\cos(\kb_0\cdot\rb^\prime) n(\tau^\prime ,\rb^\prime)\ ,
\end{align}
where $n(\tau,\rb)=\psi^\dag(\tau,\rb) \psi(\tau,\rb)$ is the atom density.
 The interaction is retarded on a scale which is set by the real part of the photon propagator
\begin{equation}
\mathcal{G}(\tau)=\frac{\cosh\left[-\Delta_{\rm c}\left(\beta/2-|\tau|\right)\right]}{\sinh\left[-\Delta_{\rm c}\beta/2\right]}\, .
\end{equation}
In the limit of a large and negative detuning $|\Delta_c|\gg\kbz T$,
$\Delta_{\rm c}<0$ 
%{\color{red} {\it WZ: compared to $k_BT$ but do we need another scale here?}}
this retardation is negligible, however, because $|\Delta_{\rm
  c}|\mathcal{G}(\tau)\to\delta(\tau)$. We recall that to arrive at Eq.~(\ref{long_range_int}), we also took 
the detuning atom-pump detuning $\Delta_a$, 
typically in the {\rm GHz} regime, to be much larger than 
the recoil energy and the atomic chemical potential, typically in the {\rm KHz} regime. In that regime, we can also safely discard potential retardation effects arising from propagation of the intermediate excited state in the 2-photon processes described 
earlier in Fig.~\ref{effective_vertices}.
% The latter disappear in the regime where the cavity detuning
% is much larger than the typical atomic time scales $\kbz T,\mu$. In
% this adiabatic cavity regime the propagator can indeed be approximated by a
% Dirac delta function i.e. the interaction is instantaneous.
The effective interaction Hamiltonian which is associated with the
action (\ref{long_range_int}) can be written in the form
\begin{equation}
\hat{H}_{\rm
  int}=-\frac{\lambda^2}{2N|\Delta_{\rm c}|}\hat{\rho}_{\kb_0}\hat{\rho}_{\kb_0}\;,
\end{equation}
where $\hat{\rho}_{\kb_0}=\int_{V}\! d\rb\ \hat{\psi}^\dag(\rb)
\hat{\psi}(\rb)\cos(\kb_0\cdot\rb)$ is the Fourier transform of the atom density 
at wave vectors $\pm\kb_0$. 

Remarkably, the above interaction term Eq.~(\ref{long_range_int}) is of the same form 
than the interaction term in the reduced BCS model of attractively
interacting fermions upon replacing
$\hat{\rho}_{\kb_0}\hat{\rho}_{\kb_0}$ with $\hat{B}_{0}^\dag
\hat{B}_{0}$, where $\hat{B}_{0}=\int_{V}\! d\rb\
\hat{\psi}_{\uparrow}(\rb)
\hat{\psi}_{\downarrow}^{\phantom{\dag}}(\rb)$ is the total number of zero momentum fermion bilinears.
In a classic paper \cite{muehlschlegel_1962}, M\"uhlschlegel has shown that such reduced 
models are solved exactly within mean-field theory (see also Ref. \onlinecite{haag_1962}). 
% Characteristic of these solutions is 
%a thermodynamic free energy density of the type Eq.~(\ref{s_mf}) with a logarithmic 
%dependence on the (time-independent and homogeneous-in-space) order parameter. 
Diagrammatically, this means that the exact free energy density consists of all connected 
``bubble'' diagrams in each of which the prefactor $1/N$ cancels with the loop integration over all 
degrees of freedom. Higher-order corrections from inserting additional vertices and closing external legs 
(for example of the type shown in Fig.~\ref{diag_interaction}), will generate additional vertices each scaling as 
1/N. Because the momentum-transfer at each vertex is restricted to exactly equal $\mathbf{k}_0$ here, or 
zero in the BCS model, these vanish. 

Note that for the DHL self-organization transition, the order parameter is a real-valued density wave 
(dual to the coherent cavity field appearing in Eq.~(\ref{s_mf})), whereas in the BCS model 
the order parameter is in general a complex-valued Cooper pairing amplitude. Condensation of 
Cooper pairs breaks a $U(1)$ symmetry because the operator $B_{0}$ may have 
an arbitrary phase, while in our case $\hat{\rho}_{\kb_0}$ 
is real and the only remaining degeneracy is the discrete 
parity, which corresponds to ordering in $+\kb_0$  or $-\kb_0$,
a $Z_2$ symmetry (see Ref.~\onlinecite{gersch06} for an explanation of an 
analogous $Z_2$-type mean-field model for commensurate charge ordering of fermions in a 
half-filled hypercubic lattice).

\section{Conclusions}
\label{sec:conclu}

This paper provided an exact solution of the interplay of Bose-Einstein condensation (BEC) and the Dicke-Hepp-Lieb (DHL)
self-organization transition of a Bose gas at arbitrary temperatures subject to a dynamical optical potential 
generated by the electromagnetic vacuum field of a high-quality optical resonator.

We showed that, at finite temperatures, the typically invoked Hilbert space truncation to only two atomic momentum 
states, that leads to a description in terms of a zero-dimensional effective Dicke model, becomes unphysical and breaks down. We developed an effective action approach that captures the dynamical multi-band structure, that the cavity generates for the atoms, as well as the backaction of the atoms on the cavity spectrum. Using this approach, we computed 
the full phase diagram at arbitrary temperature and discovered a bi-critical point, at which the atoms become 
superfluid and self-organize at the same time. We showed that the thermally excitable continuum of atomic finite 
momentum states can enhance self-organization in striking contrast to the finite-temperature solution of the Dicke model 
where a finite temperature always counteracts ferromagnetic spin order. Moreover, the thermal bath of atoms strongly broadens and overdamps the polaritonic sidebands of the cavity spectrum. 

We also gained some structural insights and argued that the cavity-Bose gas problem, 
in the absence of short-range interactions between the atoms, belongs to a class of exactly solvable, so called restricted,
mean-field models, the most prominent of which is the celebrated BCS-theory of superconductivity. 
It is remarkable that many-body cavity QED provides a natural, experimental incarnation of such models. 
We note that also the experimentally relevant non-equilibrium extensions of the model of this paper, that is,
coupling the photons and atoms to Markovian baths, will remain exactly solvable as long as 
potential loss and gain terms are operative on the single-particle level.

Our calculations were carried out in a frame rotating with the optical frequency of the driving laser. In that frame, 
the periodically driven Bose gas in a cavity can be formulated as a time-independent Hamiltonian problem. We here employed a equilibrium path integral formalism without explicitly accounting for the finite photon lifetimes and spontaneous emission of the atoms. This means we have implicitly assumed a dissipative loss channel for the energy deposited by the drive laser without 
accounting for the usually arising noise terms (from the fluctuation-dissipation theorem). This approach is not rigorous but has been shown to work well for the unitary (reversible) sector of the non-equilibrium steady states for Dicke-type models in 
cavity QED (see e.g. Ref.~\onlinecite{strack_glass,dallatorre_2012,sarang_2010} for detailed comparison between the equilibrium and non-equilibrium steady states). In particular the phase diagram and basic features of the spectral properties 
only receive relatively small, quantitative modifications from the dissipation. The fact that the Markovian baths typical of quantum optics generate a finite, effective temperature at long times further supports the usage of an equilibrium formalism. 
In the present context, the main aspect that the equilibrium formalism misses are (additional) imaginary contributions to the poles of correlation functions which should lead to irreversible, overdamped dynamics especially close to the self-organization transition. 
The equilibrium formalism also is not able to differentiate between the transient regimes at shorter times, which should be the most relevant for a refined comparison with experiments, and the asymptotic long-time regimes, which have so far been the focus 
of most of the theoretical works.

In the future, it will be interesting to generalize the results obtained in the present paper to lossy cavities and clarify further the dissipative effects induced by a finite temperature versus those induced by cavity decay. It should also be possible 
to consider interacting Bose gases in a cavity and study the competition of cavity-mediated interaction, short-range repulsion, and dissipation in the superfluid regime.

\begin{acknowledgments}

We thank Mikhail D. Lukin, Steffen P. Rath, and Richard Schmidt for
helpful discussions and gratefully acknowledge useful remarks on the manuscript by 
Michael Buchhold and Sarang Gopalakrishnan. 
F. P. acknowledges support from the Alexander Von Humboldt foundation. The research of PS was supported by the DFG under grant Str 1176/1-1, by the NSF under Grant DMR-1103860, by the Army Research Office Award W911NF-12-1-0227, by the Center for Ultracold Atoms (CUA) and by the Multidisciplinary University Research Initiative (MURI). 

\end{acknowledgments}

\appendix

\section{Derivation of the photon propagator}
\label{derivation_propagator}

Our goal in this appendix is to derive the action (\ref{S_eff_2}) from
the full effective action (\ref{S_eff}) by expanding the latter to
second order in the cavity field $a_{\n}$. To this purpose, we write
the Nambu matrix (\ref{matrix_full}) as:
\begin{equation}
\mathrm{M}_{\rm n,m}(\mathbf{k})=\delta_{\rm
  n,m}\mathrm{G_0}^{-1}(\omega_{\n};\kb)+\mathrm{Q}_{\rm n,m}\ ,
\end{equation}
where the matrix $\mathrm{G_0}^{-1}(\omega_{\n};\kb)$ is diagonal in Nambu space
with elements
$(\dots,G_0^{-1}(\omega_{\n};\kb-\kb_0),G_0^{-1}(\omega_{\n};\kb),G_0^{-1}(\omega_{\n};\kb+\kb_0),\dots)$,
and the matrix
\begin{equation}
\mathrm{Q}_{\rm n,m}=
\left(
\begin{array}{ccc}
\dots & \dots & \dots \\
U_{\rm n,m} & \Lambda_{\rm n,m}/2 &  U_{\rm n,m}/2\\
\Lambda_{\rm n,m}/2 & U_{\rm n,m} & \Lambda_{\rm n,m}/2 \\
 U_{\rm n,m}/2 & \Lambda_{\rm n,m}/2 & U_{\rm n,m} \\
\dots & \dots & \dots
\end{array} 
\right)\ .
\end{equation}
We then expand the tracelog and the inverse as follows:
\begin{align}
\mathrm{Tr}\ln\left(\mathbb{I}_{\n}\mathrm{G_0}^{-1}+\mathrm{Q}\right)=\nonumber\\\mathrm{Tr}\ln\left(\mathbb{I}_{\n}\mathrm{G_0}^{-1}\right)+\mathrm{Tr}\left(\mathbb{I}_{\n}\mathrm{G_0}\mathrm{Q}\right)-\frac12 \mathrm{Tr}\left[(\mathbb{I}_{\n}\mathrm{G_0}\mathrm{Q})^2\right]+\dots
\ ,\nonumber\\
\left(\mathbb{I}_{\n}\mathrm{G_0}^{-1}+\mathrm{Q}\right)^{-1}=\nonumber\\\mathbb{I}_{\n}\mathrm{G_0}-\mathbb{I}_{\n}\mathrm{G_0}\mathrm{Q}\mathbb{I}_{\n}\mathrm{G_0}+(\mathbb{I}_{\n}\mathrm{G_0}\mathrm{Q})^2\mathbb{I}_{\n}\mathrm{G_0}+\dots\ ,
\end{align}
where $\mathbb{I}_{\n}$ is the identity matrix in Matsubara space.
As an example, let us consider the term:
\begin{widetext}
\begin{align}
\mathrm{Tr}\ln\left(\mathbb{I}_{\n}\mathrm{G_0}^{-1}\right)&=\sum_{\rm
  n}\sum_{\kb\in
  B}\left[\dots+\ln\left(G_0^{-1}(\omega_{\n};\kb-\kb_0)\right)+\ln\left(G_0^{-1}(\omega_{\n};\kb)\right)+\ln\left(G_0^{-1}(\omega_{\n};\kb+\kb_0)\right)+\dots\right]=\nonumber\\
&=\sum_{\rm  n}\sum_{\kb}\ln\left(G_0^{-1}(\omega_{\n};\kb)\right)\ .
\end{align}
\end{widetext}
For the last equality, we have transformed a sum of terms each with a
momentum sum restricted to the first Brillouin zone into a single term
with an unrestricted momentum sum. 
With analogous manipulations we can
also obtain:
\begin{align}
\label{disp_2}
\mathrm{Tr}\left(\mathbb{I}_{\n}\mathrm{G_0}\mathrm{Q}\right)&=\frac{u_0}{2N\beta}\sum_{\rm
n}|a_{\n}|^2\sum_{\n_1,\kb}G_0(\omega_{\n_1};\kb)\nonumber\\&=\frac{u_0(1-N_0/N)}{2\beta}\sum_{\rm
n}|a_{\n}|^2\ ,
\end{align}
up to second order in $a_{\n}$.
For the last equality we have used the fact, as discussed in
section \ref{TL}, that the ideal gas equation of state is exact in the
TL whereby $\sum_{\n,\kb}G_0(\omega_{\n};\kb)=N-N_0$.
Up to second order in $a_{\n}$, the last contribution we get from the
tracelog comes from:
\begin{widetext}
\begin{align}
-\frac12
\mathrm{Tr}\left[(\mathbb{I}_{\n}\mathrm{G_0}\mathrm{Q})^2\right]=\nonumber\\-\frac{1}{2}\sum_{\rm
  n,m}\sum_{\kb\in  B}\frac{\Lambda_{\rm n,m}\Lambda_{\rm
    m,n}}{2}\left[\dots+G_0(\omega_{\rm n};\kb) G_0(\omega_{\rm
    m};\kb-\kb_0)+G_0(\omega_{\rm n};\kb) G_0(\omega_{\rm
    m};\kb+\kb_0)+\right.\nonumber\\
\left.+G_0(\omega_{\rm n};\kb-\kb_0) G_0(\omega_{\rm
    m};\kb)+G_0(\omega_{\rm n};\kb+\kb_0) G_0(\omega_{\rm
    m};\kb)+\dots\right]=\nonumber\\
-\frac{\lambda^2}{2n}\sum_{\rm
  n}\frac12\Pi(\omega_{\n};\kb_0)\left(a_{\n}^{*}a_{\n}^{\phantom{*}}+a_{-\n}^{*}a_{-\n}^{\phantom{*}}+a_{\n}^{*}a_{-\n}^{*}+a_{\n}^{\phantom{*}}a_{-\n}^{\phantom{*}}\right)\ .
\end{align}
\end{widetext}
The above contribution corresponds to the first term in the
self-energy (\ref{self_en_pol}).

Let us now turn to the expansion of the last term of the action
(\ref{S_eff}). The only second order contribution comes from the first
term in the expansion of $\mathrm{M}^{-1}$ given above. Taking into
account that the condensate spinor is simply
$\Phi_0^T=(\dots,0,\sqrt{N_0/N},0,\dots)$, this second order
contribution reads:
\begin{widetext}
\begin{align}
-N\!\!\!\sum_{\rm
  n,m\neq 0}\!\!\!\Phi_0^\dag\mathrm{M}_{\rm 0,n}(\mathbf{0}) \delta_{\rm
  n,m}\mathrm{G_0}(\omega_{\n};\kb)\mathrm{M}_{\rm
  m,0}(\mathbf{0})\Phi_0=\nonumber\\=-N_0\sum_{\rm n\neq 0}\frac{\Lambda_{\rm n,0}\Lambda_{\rm
    0,n}}{4}\left(G_0(\omega_{\rm n};-\kb_0)+G_0(\omega_{\rm
    n};\kb_0)\right)=-\frac{\lambda^2N_0/N}{2\beta}\sum_{\rm n\neq
  0}\frac{E_{\rm R}}{\omega_{\n}^2+E_{\rm R}^2}\left(a_{\n}^{*}a_{\n}^{\phantom{*}}+a_{-\n}^{*}a_{-\n}^{\phantom{*}}+a_{\n}^{*}a_{-\n}^{*}+a_{\n}^{\phantom{*}}a_{-\n}^{\phantom{*}}\right)\ .
\end{align}
\end{widetext}
For the last equality we used the fact that $\Lambda_{\rm
  n,0}\Lambda_{\rm 0,n}=\Lambda_{-\rm n,0}\Lambda_{\rm 0,-n}$ in order
to sum the propagator $G_0(\omega_{\rm n};\kb_0)$ with its complex
conjugate. The above contribution corresponds to the second term in
the self-energy (\ref{self_en_pol}). The first term on the second line in
Eq.~(\ref{S_eff}) can be written as
\begin{equation}
N\Phi_0^\dag \mathrm{M}_{\rm 0,0}(\mathbf{0})\Phi_0=-\mu\beta
N_0+\frac12 u_0(N_0/N)\frac{1}{\beta}\sum_{\n}|a_{\n}|^2\ .
\end{equation}
The last term above sums up with the Eq.~(\ref{disp_2}) to yield the
self-energy (\ref{self_en_disp}).

\end{document}